\documentclass[11pt,a4paper]{article}
\usepackage{multicol}
\usepackage{comment}
\usepackage{booktabs}
\usepackage{axodraw4j}
\usepackage{url}
\usepackage[dvipsnames]{xcolor}
\usepackage{jheppub}
\usepackage{caption, subcaption}  
\usepackage{color,latexsym, array,multirow,  verbatim, enumerate,cancel,graphicx}
\usepackage{slashed}
\usepackage{bm}

\def\beq {\begin{equation}}
\def\eeq {\end{equation}}
\def\bi {\begin{itemize}}
\def\ei {\end{itemize}}
\def\bea {\begin{eqnarray}}
\def\eea {\end{eqnarray}}

\numberwithin{equation}{section} 
\title{Search for Electroweakinos in R-Parity Violating SUSY with Long-Lived Particles at HL-LHC}

\author{Biplob Bhattacherjee$^1$, Prabhat Solanki$^1$}

\affiliation{\vspace*{0.1in}$^1$ Centre for High Energy Physics, Indian Institute of Science, Bengaluru 560012, India}

\emailAdd{biplob@iisc.ac.in}
\emailAdd{prabhats@iisc.ac.in}

\abstract{ We investigate the R-parity violating (RPV) supersymmetric (SUSY) model at the High-Luminosity Large Hadron Collider (HL-LHC) in the context of compact muon solenoid (CMS) experiment assuming a total integrated luminosity of $\mathcal{L}=3000~\text{fb}^{-1}$ at $\sqrt{s}=$ 14 TeV. We focus on the pair production of electroweakinos, specifically, $\chi_2^0$ and $\chi_1^{\pm}$ in wino and higgsino states in a particular scenario where $\chi_2^0$ and $\chi_1^{\pm}$ decay into a Higgs boson and W boson, respectively, along the long-lived lightest supersymmetric particle (LSP), $\chi_1^0$, which decays to three quarks via $\lambda^{''}$ RPV couplings leading to the prompt as well as displaced signatures in the final state. To select events at the level-1 (L1) trigger system, we employ dedicated and standard triggers followed by an offline analysis integrating information from the tracker, electromagnetic calorimeter (ECAL) and minimum ionising particle (MIP) timing detector (MTD). We observe that wino-like $\chi_2^0/\chi_1^{\pm}$ with a mass of 1900 GeV and $\chi_1^0$ with a mass greater than 800 GeV can be probed across a decay length ranging from 1 cm to 200 cm. In the case of higgsino-like pair production of $\chi_2^0/\chi_1^{\pm}$, we can probe $\chi_2^0/\chi_1^{\pm}$ with a mass of 1600 GeV, and $\chi_1^0$ with a mass greater than 700 GeV, across a decay length range of 1 cm to 200 cm.
}
 
\begin{document}
\maketitle
\flushbottom


\section{Introduction}
\label{sec:intro}

With a growing and urgent need to search for physics beyond the standard model (BSM), there is an ongoing effort to look for signatures of new physics in the long-lived sector on both the phenomenological and experimental sides. Numerous phenomenological studies focusing on a wide range of BSM models and signatures have been performed to search for long-lived particles (LLPs); references to some of these studies can be found here \cite{Ilten:2015hya,Gago:2015vma,Banerjee:2017hmw,KumarBarman:2018hla,Bhattacherjee:2019fpt,CidVidal:2019urm,Banerjee:2019ktv,Jones-Perez:2019plk,Bhattacherjee:2020nno,Bhattacherjee:2021qaa,Adhikary:2022goh,Gershtein:2020mwi,Fuchs:2020cmm,Cheung:2020ndx,Liu:2020vur,Evans:2020aqs,Linthorne:2021oiz,Alimena:2021mdu,Bhattacherjee:2021rml,Sakurai:2021ipp,Du:2021cmt}. On the experimental side, LHC's two general-purpose detectors, ATLAS and CMS, have been actively searching for displaced signatures at the colliders. Studies done at ATLAS and CMS look for a wide range of experimental signatures using vertex and non-vertex-based methods. For vertex-based searches, signatures include displaced jets, vertices, and leptons. On the other hand, non-vertex-based searches feature signatures such as 
emerging jets, trackless jets, disappearing tracks, non-pointing photons, and jets with low electromagnetic energy fraction. CMS \cite{CMS:2014hka,CMS-PAS-EXO-16-022,CMS-PAS-EXO-16-036,CMS:2017kku,CMS:2018lab,CMS:2018bvr,CMS-PAS-FTR-18-002,CMS-PAS-EXO-19-013,CMS:2019qjk,CMS:2019zxa,CMS:2020atg,CMS:2020iwv,CMS:2021juv,CMS:2021kdm,CMS:2021yhb} and ATLAS \cite{ATLAS:2015xit,ATLAS:2015wsk,ATLAS:2015itk,ATLAS:2017tny,ATLAS:2018rjc,ATLAS:2018lob,ATLAS:2018niw,ATLAS:2018tup,ATLAS:2019qrr,ATLAS:2019kpx,ATLAS:2019fwx,ATLAS:2019tkk,ATLAS:2019jcm,ATL-PHYS-PUB-2019-002,ATLAS:2020xyo,ATLAS-CONF-2021-015,ATLAS-CONF-2021-032,ATLAS:2020wjh,ATLAS:2021jig,ATLAS:2023oti} have extensively documented these studies. Along with CMS and ATLAS, LHCb has also carried out numerous LLP searches involving displaced jets, dark photons and displaced leptons \cite{LHCb:2016buh,LHCb:2016inz,LHCb:2016awg,LHCb:2017xxn,LHCb:2019vmc,LHCb:2020akw}. Significant efforts are also being intensively made in the field of hardware development to improve the detection of LLPs at the large lifetime frontier. This includes the development of new detectors like FASER \cite{Feng:2017uoz},  MATHUSLA \cite{Curtin:2018mvb,Alpigiani:2020iam} and CODEX-b \cite{Gligorov:2017nwh,Aielli:2019ivi}, and hardware specifically designed for search of displaced physics at the LHC's general purpose detectors along with application of innovative analysis techniques that utilise a variety of information from the different sub-detectors at HL-LHC. There are several proposals for dedicated detectors for LLP searches at future colliders like FCC-ee \cite{Chrzaszcz:2020emg,Schafer:2022shi}. For FCC-hh, a transverse detector, DELIGHT \cite{Bhattacherjee:2021rml}, and a forward detector, FOREHUNT \cite{Bhattacherjee:2023plj} have been proposed. 

In the context of LLPs, Supersymmetry (SUSY) \cite{WESS197439,NILLES19841,HABER198575} has been one of the most studied BSM theory. In MSSM, conservation of R-parity \cite{Barbier:2004ez} ensures that SUSY particles can be pair-produced with odd R-parity where decay of each of them should lead to an odd number of sparticles and lightest SUSY particle (LSP) is stable. Numerous studies have investigated the phenomenological implications of R-parity conserving (RPC) scenarios \cite{Hooper:2002nq, Bhattacharyya:2011se,AlbornozVasquez:2011yq,Choudhury:2012tc,Fowlie:2013oua,Roszkowski:2014iqa,Hamaguchi:2015rxa,Han:2014nba,Belanger:2013pna,Ananthanarayan:2013fga,Dreiner:2012ex,Choudhury:2013jpa,Calibbi:2011ug,Belanger:2003wb,Belanger:2000tg,Barman:2017swy,Chakraborti:2017dpu}. Although, R-parity conservation is required to avoid unwanted B- and L- violating effects \cite{Hinchliffe:1992ad, Bhattacharyya:1998dt} such as proton decay but it is not the utmost requirement as presence of some other symmetries can allow R-parity violation (RPV) while forbidding proton decay \cite{Ibanez:1991hv,Dreiner:2005rd,Dreiner:2006xw,Dreiner:2012ae}. All standard model particles are assigned $R_p$ = 1 while their super-partners are assigned $R_p$ = -1.
 
 A viable MSSM superpotential comprising of gauge invariant and R-parity violating terms \cite{Martin:1997ns,Mohapatra:2015fua} can be constructed as follows- 
 
\begin{equation}
    W = \mu_iH_uL_i + \frac{1}{2}\lambda_{ijk}L_iL_jE_k^c + \frac{1}{2}\lambda_{ijk}^{'}L_iQ_jD_k^c + \frac{1}{2}\lambda_{ijk}^{''}U_i^c D_j^c D_k^c
\end{equation}

Where $\lambda_{ijk}$, $\lambda_{ijk}^{'}$ and $\lambda_{ijk}^{''}$ are various RPV yukawa couplings with i, j, k being generation indices. E, U and D represents the superfields for right-handed lepton, up-type quark and down-type quark respectively while L and Q corresponds to left-handed lepton and quark superfields respectively while $H_u$ represents superfield for up-type Higgs. In current study, we only focus on $\lambda_{ijk}^{''}$ yukawa coupling where a sparticle decays to quarks with very small coupling leading to the sparticle being long-lived. Examples of such LLPs in context of MSSM can be electroweakinos ($\chi_1^{\pm}$, $\chi_2^{0}$ and $\chi_1^{0}$ ) and gluinos. Various experiments have set an upper limit on RPV couplings to be very small leading to SUSY particles being produced with longer lifetimes. In \cite{Goity:1994dq}, where bounds on  $\lambda_{ijk}^{''}$ are calculated from double nucleon decay into two kaons, $\lambda_{112}^{''}$ less than $10^{-15} R^{-5/2}$ is excluded where R represents ratio between hadronic and supersymmetric scales and can vary from $10^{-3}$ to $10^{-6}$. Accordingly, $\lambda_{112}^{''}$ can vary from value as low as $10^{-7}$ to 1.
Indirect bound on  $\lambda_{113}^{''}$ comes from neutron oscillations where $\lambda_{113}^{''}$ less than $10^{-4}$ is excluded for $m_{\Tilde{q}} =$ 100 GeV \cite{ZWIRNER1983103}.

 Several displaced jets searches have been performed at CMS and ATLAS to specifically set exclusion limits on the mass, lifetime and production cross-section of LLPs decaying to jets assuming different SUSY models. CMS has conducted studies on RPC SUSY scenarios involving LLPs, setting constraints on their production. Detailed results and models are elaborated in the reference \cite{CMS:2020iwv}.
 
 For RPV SUSY model where gluinos are pair-produced with each gluino decaying to a top, bottom and strange quark through $\lambda_{323}^{''}$ type UDD coupling, CMS rules out gluino pair production cross-section exceeding 0.1 fb when $c\tau$ ranges between 3 and 1490 mm and $m_{\Tilde{g}}$ is 2400 GeV. Between $c\tau$ 3 mm and 1000 mm, gluinos up to 2500 GeV mass are excluded \cite{CMS:2020iwv}. CMS also studies two other RPV models where top squarks are pair-produced and each squark subsequently decays to a lepton and a bottom or a down type quark via $\lambda_{x33}^{'}$ or $\lambda_{x13}^{'}$ LQD type RPV coupling. For RPV model with $\lambda_{x13}^{'}$ LQD type coupling, production cross-sections of stop above 0.1 fb are excluded for $c\tau$ between 8 mm and 160 mm for  $m_{\Tilde{t}}=$ 1600 GeV. For $c\tau$ between 5 mm and 240 mm, top squark masses up to 1600 GeV are excluded \cite{CMS:2020iwv}. 
 
 For RPV model with $\lambda_{x33}^{'}$ LQD type coupling, stop production cross-sections exceeding 0.1 fb are excluded for $c\tau$ between 7 mm and 220 mm for  $m_{\Tilde{t}}=$ 1600 GeV. Top squark masses up to 1600 GeV are excluded for $c\tau$ between 3 mm and 360 mm \cite{CMS:2020iwv}. Another study done at CMS to study nonholomorphic RPV coupling where top squarks undergo pair-production and then decay to two down type anti-quarks each, for a top squark mass $m_{\Tilde{t}} = $ 1600 GeV, the production cross sections exceeding 0.1 fb are ruled out for $c\tau$ ranging between 3 mm and 820 mm. Additionally, for $c\tau$ values between 2 mm and 1320 mm, top squark masses up to 1600 GeV are excluded \cite{CMS:2020iwv}.

 A recent study performed at the ATLAS experiment has set up very stringent exclusion limits on masses of displaced electroweakinos in two benchmark scenarios of LLPs decaying to jets via UDD-type RPV coupling \cite{ATLAS:2023oti}. In the first scenario, electroweakinos are pair-produced in pure higgsino state which includes four possible combinations of electroweakinos: $\chi_1^{\pm}\chi_2^{0}$, $\chi_2^{0}\chi_1^{0}$, $\chi_1^{+}\chi_1^{-}$ and $\chi_1^{\pm}\chi_1^{0}$ while the other scenario involves pair-production of gluinos  ($\Tilde{g}$)  where each gluino decays promptly to a long-lived neutralino and a quark and anti-quark pair with 100\% branching ratio. In each scenario, electroweakinos decay to light flavour quarks via the $\lambda^{''}$ coupling with 100\% branching ratio. Electroweakinos with masses less than 1500 GeV are excluded for mean proper lifetime between 0.03 ns ($c\tau=$ 0.9 cm) to 1 ns ($c\tau=$ 30 cm) for pair-produced electroweakinos, while electroweakinos with masses less than 1500 GeV are excluded for mean proper lifetime between 0.02 ns ($c\tau=$ 0.6 cm) to 4 ns ($c\tau=$ 120 cm) for electroweakinos produced through the decay of gluinos with a mass of 2.4 TeV. In the context of the present analysis concerning pair-produced electroweakinos, we observe weaker limits as we increase the decay length of the LLPs above 30 cm. 

In conclusion, based on the displaced searches performed at both CMS and ATLAS, we observe that exclusion limits set for the masses of displaced gluinos are significantly high, with gluinos having masses up to 2.5 TeV already excluded at CMS \cite{CMS:2020iwv}. However, the limits imposed on the masses of displaced electroweakinos are moderate and can still be probed at future colliders like HL-LHC \cite{Apollinari:2284929}. It is also important to highlight that while we observe stronger limits for LLPs with smaller lifetimes, the limits placed on the electroweakinos are considerably lenient for highly displaced ones. For example, in the scenario described in \cite{ATLAS:2023oti}, where electroweakinos are pair-produced with a decay length of about 500 cm, the excluded electroweakino mass reduces from 1500 GeV to roughly 1050 GeV.

In this paper, we exclusively focus on the CMS detector at HL-LHC, one of the general-purpose detectors that will undergo several major hardware and software upgrades. At the HL-LHC, the peak instantaneous luminosity is set to rise to \(5 \times 10^{34}\) (\(7.5 \times 10^{34}\)) \(\mathrm{cm^{-2} s^{-1}}\), with each \(pp\) collision witnessing 140 (200) pile-up interactions. HL-LHC is projected to record data corresponding to an integrated luminosity of 3000 (4000) $\mathrm{fb^{-1}}$ during its lifetime.
In order to deal with increased PU interactions and maintain the optimal physics performance of the detectors, several hardware upgrades will take place starting with the upgrade of trigger and data acquisition systems (DAQ). With the upgrade of both the inner and outer tracker and the implementation of Field Programmable Gate Arrays (FPGA), there will be a significant overhaul in the data acquisition process at level-1 (L1) of the trigger system \cite{CERN-LHCC-2020-004}. This upgrade enables the availability of tracking information at L1. Additionally, calorimeter information from ECAL and HCAL will also be made available at L1 \cite{CERN-LHCC-2020-004}. The improved data acquisition and processing architecture at L1 will make it possible to implement particle flow and machine learning techniques, along with higher-level object reconstruction, to be used in the trigger system. This will be immensely helpful in recording rare BSM events, such as events containing displaced objects, that would have otherwise gone unrecorded. The implementation of extended tracking at L1 will enable the reconstruction of displaced tracks up to a certain transverse impact parameter which will again be very helpful in selecting events with displaced signatures at L1. Displaced particle searches will also benefit from the availability of timing information from the upgraded ECAL at L1 \cite{CERN-LHCC-2020-004, CERN-LHCC-2017-011} and the inclusion of an all-new MIP timing detector (MTD) between the tracker and calorimeter system \cite{CMS:2667167}. Additionally, a new high granularity calorimeter (HGCAL) will replace the existing endcap calorimeter \cite{CERN-LHCC-2017-023}, enhancing the physics performance in the forward region under the harsher conditions at the HL-LHC.

The upgrades planned for the HL-LHC will substantially boost physics sensitivity and increase the probing potential of LLPs at HL-LHC. However, there are not many comprehensive and realistic phenomenological studies explicitly designed for HL-LHC that fully consider the effect of increased PU and make the most of the impending hardware upgrades at HL-LHC. This motivates us to investigate the lifetime frontier of BSM physics in the context of RPV SUSY, considering increased PU conditions at HL-LHC.

Rest of the paper is structured as follows: In Section \ref{sec:sigback}, we outline the signal model, background sources, and the simulation setup for both signal and background events. Section \ref{sec:trigger} explains the implemented triggering strategy at L1, where we select events at L1 by utilizing available information from the upgraded detector systems. In Section \ref{sec:offana}, we perform a detailed analysis of the events selected at L1. This analysis involves studying various physics variables constructed using offline information from different sub-detectors at the CMS. The analysis is divided into three parts: a cut-based analysis and two independent multi-variate analyses. Section \ref{sec:results} presents the signal significance for various LLP benchmark points, providing quantitative results for our analysis. Finally, in Section \ref{sec:summary}, we summarize and draw conclusions based on our analysis.

\section{Signal Model, Backgrounds, and Simulation Setup}
\label{sec:sigback}

\begin{figure}[hbt!]
    \centering
    \includegraphics[scale=0.5]{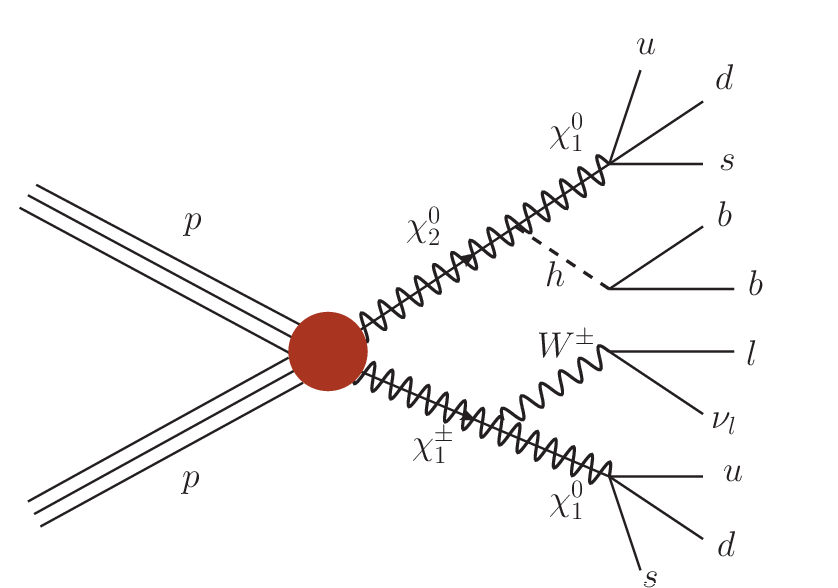} ~~
    \caption{Feynman diagram of cascade decay of electroweakinos ($\chi_{2}^{0}/\chi_{1}^{\pm}$) where $\chi_{2}^{0}$ decays to $\chi_1^0$ and 125 GeV Higgs boson while $\chi_1^{\pm}$ decays to a W boson and $\chi_1^0$.  }
    \label{fig:chi2chi1_feyn}
\end{figure}

In this paper, we study R-parity violating yukawa coupling of type $\lambda^{''}$ within the framework of MSSM. Our focus is on the associated production of electroweakinos, specifically the $\chi_{2}^{0}$ and $\chi_{1}^{\pm}$ where $\chi_{2}^{0}$ decays to lightest supersymmteric particle (LSP), $\chi_1^0$, and the 125 GeV Higgs boson while the $\chi_1^{\pm}$ decays to a W-boson and $\chi_1^0$.  Due to a very small $\lambda^{''}$ coupling, the $\chi_1^0$ exhibits a long lifetime. We consider decay of $\chi_1^0$ to light flavor quarks (u, d, and s) with 100\% branching ratio.  We assume a 100\% branching fraction for the decays $\chi_{2}^{0} \rightarrow \chi_1^0,h$ and $\chi_{1}^{\pm} \rightarrow \chi_1^0,W^{\pm}$. The inclusive decays of the Higgs boson and W boson are considered, with their respective branching ratios taken from  \cite{ParticleDataGroup:2022pth}.  Quarks resulting from decay of $\chi_1^0$ undergo showering and hadronization leading to the production of multiple displaced jets in the final state. Feynman diagram illustrating the cascaded decay process assuming only one decay mode for both Higgs boson and W boson is shown in Figure \ref{fig:chi2chi1_feyn}. In this diagram, the Higgs boson decays exclusively into two b-jets, while the W boson decays into leptons.

The pair production cross-section of the neutralino-chargino pair at $\sqrt{s}=$ 14 TeV is calculated at the next-to-leading order (NLO) with the incorporation of next-to-leading-log (NLL) effects, using the \texttt{RESUMMINO} code  \cite{Fiaschi:2023tkq}. For the current analysis, we solely focus on the pair production of electroweakinos with degenerate masses ($m_{\chi_{2}^{0}} = m_{\chi_{1}^{\pm}}$). 
The SUSY cross-sections for electroweakino pair production provided by the LHC collaboration \cite{cms:cross} matches with those we got from \texttt{RESUMMINO}\footnote{\texttt{RESUMMINO} gives a cross-section of 0.124 and 0.051 $\mathrm{fb}$ for pair production of 1500 GeV wino-like and 1400 GeV higgsino-like electroweakino masses.}.

 We study LLPs, $\chi_1^{0}$, in a mass range varying from 500 GeV to 1 TeV, with mean proper decay length varying from 1 cm to 500 cm. For signal generation, as well as for showering and hadronization, we utilize $\texttt{pythia8}$ \cite{Sjostrand:2014zea}. The signal samples are generated using the $\texttt{CTEQ6L1}$ PDF (Parton Distribution Function) \cite{DanielStump_2003}  with the $\texttt{CUETP8S1-CTEQ6L1}$ CMS tune \cite{CMS:2015wcf}.  During the sample generation, we adjust the decay width of the LLPs in the input SLHA (Supersymmetry Les Houches Accord) file to modify the decay length of the LLPs.

Since our signal signature includes multiple jets in the final state, the main source of background arises from the instrumental effects and QCD multijet processes. Additionally, due to the presence of leptons in the final state, a sub-dominant contribution to the background comes from $t\bar{t}$ events in which the top quark can decay leptonically or hadronically. We also anticipate background contribution from W+jets events, where the W boson decays inclusively. Here, we would like to mention that simulating and characterising the instrumental effects are out of the scope of the current study. Instead, we focus on the mitigation strategy for the instrumental background outlined in the subsequent sections.  

QCD events are generated in the bins of parton level $H_T$ ($H_T^{gen}$). Here, $H_T^{gen}$ is calculated by summing the transverse momenta of all partons involved in the event. $H_T^{gen}$ bins used in this study include following ranges - 500 - 600 GeV, 600 - 700 GeV, 700 - 800 GeV, 800 - 1000 GeV and $>$1000 GeV. $H_T^{gen}$ bins are selected based on the analysis
strategy, where events in the offline stage of the study after being triggered at level-1 by the triggers, as elaborated in Section \ref{sec:trigger}, are required to have high event $H_T$. This is because signal events can easily surpass $H_T>$ 500 GeV threshold due to significant hadronic activity in the final state. Therefore, QCD multijet events are generated in $H_T^{gen}$ bins starting with $H_T^{gen} >$ 500 GeV in order to ensure sufficient background statistics. Generation of background events is done in \texttt{madgraph} \cite{Alwall:2014hca,Frederix:2018nkq} while showering is done using \texttt{pythia8}.

We use \texttt{Delphes-3.5.0} \cite{deFavereau:2013fsa} for simple detector simulation. To accurately replicate the conditions at HL-LHC, which are characterized by a high PU environment, our analysis takes into account the effects of PU. PU originates from the multiple soft proton-proton interactions that occur within a single bunch crossing, along with a hard collision. We use \texttt{PYTHIA8} to generate 1 million soft QCD events which are utilized as PU events. The \texttt{PileUpMerger} module in \texttt{Delphes} subsequently merges these PU events with the hard process. Both signal and background events have an average of 140 PU events.

We use default CMS card provided with \texttt{Delphes} for HL-LHC for detector simulation. However, we make specific modifications to certain \texttt{Delphes} modules as elaborated in one of our previous studies \cite{Bhattacherjee:2021qaa}. To form jets using energy deposits from the calorimeters, ECAL and HCAL, we use the anti-$k_T$ jet clustering algorithm \cite{Cacciari:2008gp} with a cone size of R = 0.3. Using a narrower jet cone size instead of the standard R = 0.4 was motivated by the need to mitigate contamination from PU interactions, which can significantly affect the measurement of physics variables. The amount of PU contamination within a jet relies on the jet area, as PU is distributed throughout the detector. A reduction in the jet area leads to smaller PU contribution. By shrinking the jet cone size, the effects of PU can be effectively reduced, assuming that the jets from the signal process remain unaffected, and that the majority of the hadronic activity from the signal is captured within a reduced cone radius. This approach aligns with our analysis, as prior studies \cite{Banerjee:2017hmw, Bhattacherjee:2019fpt, Bhattacherjee:2020nno, Bhattacherjee:2021qaa} have shown that displaced jets resulting from LLP decays typically concentrate energy within a more confined region of the $\eta - \phi$ plane. Consequently, opting for a narrow cone size for jets can aid in minimizing the impact of PU on LLP jets. 

Before we proceed further, let's define two signal benchmark points (BP) for our analysis:

\begin{itemize}
\item \textbf{BP-1}: $M_{\chi_2^{0}}/M_{\chi_1^{\pm}}$ = 1600 GeV, $M_{\chi_1^{0}}$ = 800 GeV, and $c\tau$ = 10 cm.
\item \textbf{BP-2}: $M_{\chi_2^{0}}/M_{\chi_1^{\pm}}$ = 1600 GeV, $M_{\chi_1^{0}}$ = 800 GeV, and $c\tau$ = 100 cm.
\end{itemize}

Here, $c\tau$ represents the mean proper decay length of the LLP. We have selected benchmark points (BP-1 and BP-2), keeping in mind the stringent limit on the masses of electroweakinos. Both BP-1 and BP-2 feature moderately heavy LLPs resulting from the decay of significantly heavy electroweakinos, $M_{\chi_2^{0}}/M_{\chi_1^{\pm}}=$ 1600 GeV. These electroweakinos have an extremely small pair-production cross-section. We are examining two decay length scenarios: one involves a shorter decay length of 10 cm, and the other features a considerably longer decay length of 100 cm, for which the limits are still lenient. Throughout the rest of the paper, we will use the aforementioned shorthand notation to refer to the signal benchmark points. QCD events with $H_T^{gen} \in$ \{500, 600\} GeV will be represented as ``QCD," and top quark pair events will be denoted as ``$t\bar{t}$''. We generate 5 million $t\bar{t}$ events, 3 million QCD dijet events spread across the mentioned $H_T^{gen}$ bins, and 0.6 million W+jets events. For each signal benchmark point, 0.5 million events are generated. The generation and analysis of large background datasets involving 140 PU interactions present a significant challenge. The size of the simulated events using \texttt{Delphes}, including only tracks, towers, and jet branches, can reach up to 15 GB for 5000 events. This makes it impractical to produce extensive background datasets that surpass what we have already generated due to our computational limitations.

\section{Triggering LLP events at L1}
\label{sec:trigger}

The CMS experiment employs a two-level trigger system, consisting of the Level-1 (L1) and the High-Level Trigger (HLT), to identify and select interesting events for offline analysis. The HLT is a software-based trigger, while the L1 trigger is a hardware-based system with an extremely short latency period that determines the time window within which the decision to record an event is made. Because of this low latency period, performing complex physics calculations and constructing high-level physics objects using information from multiple sub-detectors can be challenging and inefficient. However, with the proposed upgrades to the data acquisition system, it becomes possible to reconstruct certain high-level physics objects and apply machine-learning (ML) techniques at the L1 trigger stage in the context of the HL-LHC. (For more comprehensive information about implementation of ML algorithms at FPGAs, please refer to \cite{Govorkova:2021utb,Neu:2023sfh} and references therein). These upgrades will involve increasing the latency period and enhancing the data bandwidth, measured in terms of event rate. These improvements will enable the design of triggers aimed explicitly at searching for LLPs. Therefore, it is crucial to efficiently utilise available resources to select events at L1 that do not overlook exotic LLP events, which typically have a very small cross-section. The final state signature for our study consists of displaced jets and prompt leptons. Our primary focus will be on triggering events using dedicated triggers to detect events containing these specific physics objects.
 
At the HL-LHC, CMS has proposed two dedicated triggers explicitly designed to select events with a displaced jets signature  \cite{CERN-LHCC-2020-004}. In addition to these dedicated LLP triggers, single-lepton triggers can further maximize the trigger efficiency \cite{Adhikary:2022goh}. We will explain the triggers used in our analysis in detail below-

\begin{itemize}
    \item \textbf{Track-}$\bm{H_T}$: At the HL-LHC, CMS plans to upgrade the inner tracker by replacing both the pixel and strip tracking detectors with smaller pixel sensors. The outer tracker will also be improved by incorporating strip and macro pixel sensors with stacked strip modules. The main requirements for the upgraded tracker system at HL-LHC include high radiation tolerance, increased granularity, improved track separation, availability of tracking information at L1, and extended tracking acceptance. The upgraded outer tracker will facilitate the reconstruction of track candidates at L1, operating at a rate of 40 MHz, for $|\eta|<$2.4. This will be achieved through an increased latency period and the implementation of FPGAs, enabling the construction of track-based triggers at L1. The availability of tracking information at L1 will warrant the identification of the primary vertex and will be immensely useful in mitigating charged PU. In addition to the advantages mentioned above of including tracking information at L1, one particular advantage relevant to this analysis is the extension of the L1 tracking algorithm to reconstruct tracks displaced within the detector. Our analysis considers a track displaced from the beamline if it has transverse impact parameter ($|d_0|$) greater than 1.5 mm. These tracks may originate from a secondary vertex following the decay of an LLP. The efficiency of track reconstruction for displaced tracks at L1 will depend on the $|d_0|$ of the tracks, with efficiency decreasing as $|d_0|$ increases. Tracking at L1 will be available for particles with transverse momentum ($p_T$) greater than 2 GeV within the pseudorapidity ($|\eta|$) range of less than two. It will follow a track reconstruction efficiency curve as shown in the reference \cite{CERN-LHCC-2020-004}. 
    
    To highlight the importance of displaced tracking at L1, Figure \ref{fig:disp_track} illustrates the L1 displaced track multiplicity within a $\Delta R <$ 0.3 cone around the jet axis for two LLP benchmark scenarios: BP-1 (decay length of 10 cm) and BP-2 (decay length of 100 cm), with $M_{\chi_1^0} = 800$ GeV and $M_{\chi_2^{0}} = 1600$ GeV for jets with $p_T>$ 40 GeV and $|\eta|<$ 2.5. The figure also shows the displaced track multiplicity for two primary background sources: $t\bar{t}$ and QCD dijet events. 
    
    \begin{figure}[hbt!]
    \centering
    \includegraphics[scale=0.5]{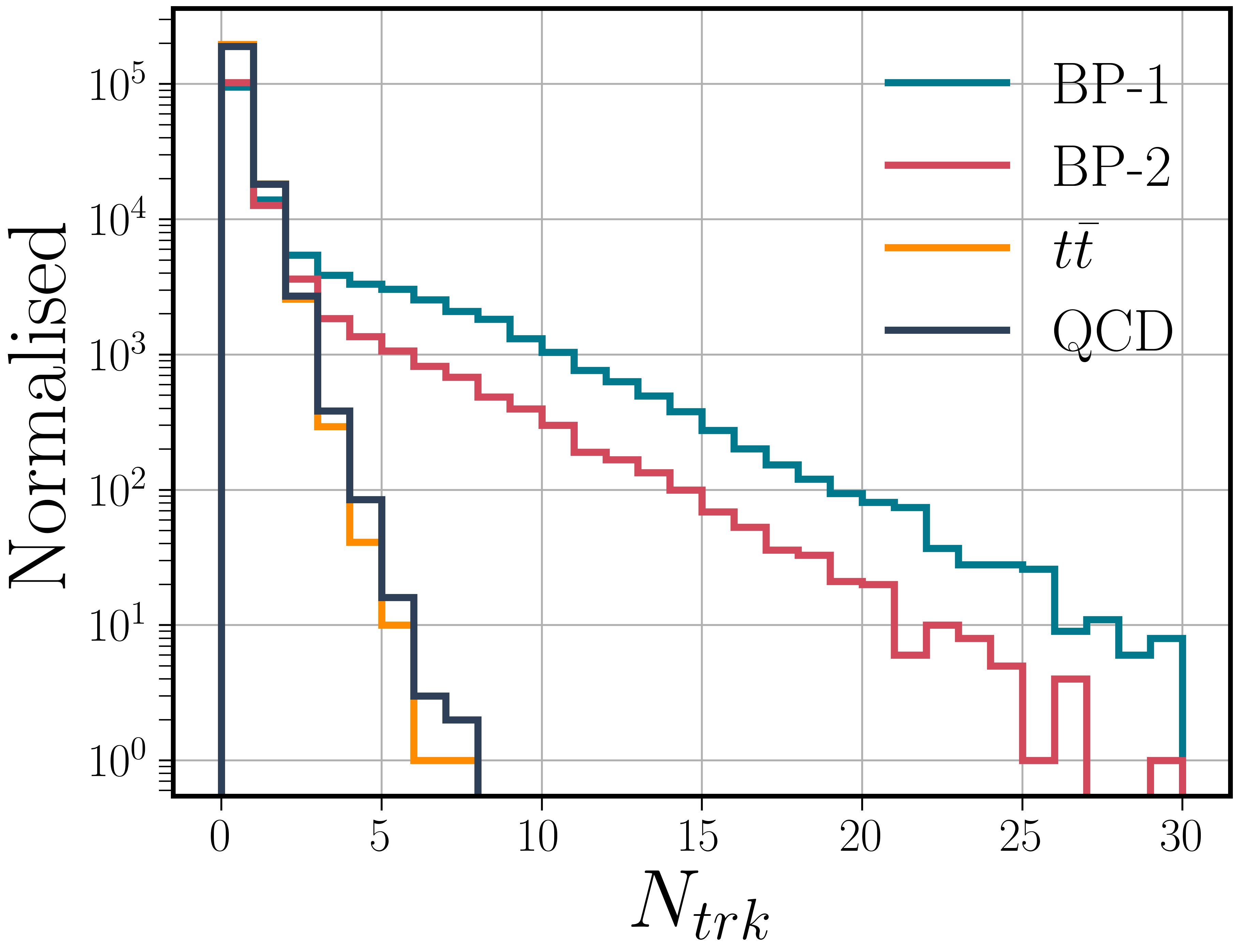} ~~
    \caption{Comparison of displaced track multiplicity within $\Delta R<$0.3 of jet for BP-1 (10 cm) and BP-2 (100 cm) LLP benchmarks, with $M_{\chi_1^0} = 800$ GeV and $M_{\chi_2^{0}}$ = 1600 GeV, along with $t\bar{t}$ and QCD background for jets with $p_T>$ 40 GeV and $|\eta|<$ 2.5}.
    \label{fig:disp_track}
    \end{figure}
    
    The figure underlines the importance of displaced tracks within jets for distinguishing between long-lived signal and background events, as we observe that the displaced track multiplicity is significantly lower for the backgrounds compared to the signal benchmark points. Moreover, the LLP benchmark (BP-1) with a shorter mean proper decay length of 10 cm exhibits a higher number of reconstructed displaced tracks which is evident from the longer tail observed in the multiplicity distribution compared to the benchmark with a longer decay length (BP-2). This observation aligns with our expectations, as LLPs with longer decay lengths will have larger value of $|d_0|$ and, therefore, fewer displaced tracks will be reconstructed.
    
    CMS has proposed a dedicated trigger for LLPs called ``Track-$H_T$" to identify events with displaced jets originating from LLPs using the upgraded tracker's improved tracking capabilities in making triggering decisions using the tracking information at L1. This trigger is specifically designed to get a handle on the events with LLPs exhibiting shorter decay lengths. The current analysis uses a track-based trigger influenced by the CMS Track-$H_T$ trigger \cite{CMS-PAS-FTR-18-018}. The Track-$H_T$ trigger used in the current study selects events where at least one displaced jet is present, and it works by calculating $H_T$ from track-based jets. 
    To form track-based jets, we begin by grouping tracks with a $p_T$ greater than 2 GeV within a $|\eta| < 2$ range. These tracks are then binned based on their closest approach to the beam line in the z-direction, $z_0$, with a bin size of 6 cm. The $z_0$ bin is chosen based on the highest scalar sum of $p_T$ of tracks. Subsequently, the tracks in the chosen $z_0$ bin are clustered into jets using the anti-$k_t$ algorithm with a cone radius of R = 0.3 for each event. Jets with $p_T >$ 5 GeV are considered for further analysis. Jets with at least two displaced tracks ($|d_0| >$ 1.5 mm) as constituents are classified as displaced jets. For events that contain at least one displaced jet in the collection, $H_T$ is calculated by summing the $p_T$ of all the jets, including those classified as displaced. In our study, an event must have a track-based $H_T$ threshold greater than 160 GeV to trigger, as inferred from \cite{CMS-PAS-FTR-18-018}.

   \item \textbf{Displaced Calo-Jet}: The upgraded ECAL at HL-LHC will provide precise timing information for ECAL energy deposits, with a timing resolution of approximately 30 ps for a 20 GeV energy deposit during the initial runs of HL-LHC \cite{CERN-LHCC-2017-011}. However, it is important to note that timing resolution may degrade over time as more data is collected. To utilize this timing information at the L1 trigger level and trigger events with displaced jets, the CMS experiment has proposed an L1 trigger incorporating ECAL timing information. For the current analysis, we utilise the L1 trigger developed in \cite{Bhattacherjee:2021qaa} that uses ECAL timing information for identifying displaced jets. 
   
  For the trigger, energy deposits from ECAL and HCAL are clustered to form jets within the $|\eta| < $ 1.5 region, utilizing the anti-$k_T$ algorithm with a cone size of R = 0.3. Each ECAL tower is required to have an energy deposit of at least 0.5 GeV, while each HCAL tower needs an energy deposit of at least 1 GeV \cite{CERN-LHCC-2020-004}. The clustering of jets is done using inputs from both ECAL and HCAL, but only the ECAL inputs are used to determine the timing of the jet. A jet is selected if at least one of the ECAL towers in its constituents has an energy deposit greater than 1 GeV \cite{CERN-LHCC-2020-004}. Each ECAL tower's timing is calibrated relative to the origin. The jet's timing is determined using the energy-weighted average of the timings from the ECAL towers inside that jet. Figure \ref{fig:ecal_time} shows the energy-weighted mean timing of jets with $p_T>$ 40 GeV and $|\eta|<$ 1.5 for two LLP benchmark scenarios, BP-1 and BP-2, along with the two main background sources.

    \begin{figure}[hbt!]
    \centering
    \includegraphics[scale=0.5]{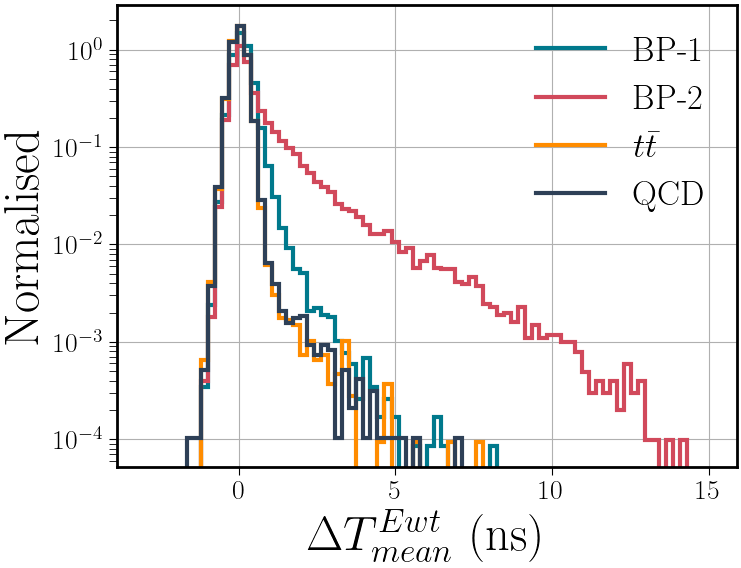}~~
    \caption{Energy weighted mean timing of jets for BP-1 and BP-2, with $M_{\chi_1^0} = 800$ GeV and $M_{\chi_2^{0}}$ = 1600 GeV, along with $t\bar{t}$ and QCD background for jets with $p_T>$ 40 GeV and $|\eta|<$ 1.5.}
    \label{fig:ecal_time}
    \end{figure}

      Figure \ref{fig:ecal_time} shows that LLPs in the benchmark scenario BP-2, characterized by longer decay lengths, exhibit higher timing values than BP-1 with shorter decay length. Furthermore, LLPs in BP-2 demonstrate significantly higher timing values than the background. For our current study, we select an event at L1 if it contains at least one jet with a timing value ($\Delta T_{\text{mean}}^{\text{Ewt}}$) greater than 1.2 ns, a jet transverse momentum ($p_T^{\text{jet}}$) greater than 35 GeV, and at least 4 ECAL towers in the jet. The threshold values used in our study are determined through explicit rate calculations as described in \cite{Bhattacherjee:2021qaa}. These calculations consider the background rate constraint for the specific scenario of 200 PU with the timing resolution at the integrated luminosity of 1000 $\mathrm{fb^{-1}}$.

    \item \textbf{Single TkIsoElectron}- Requires at least one prompt, isolated electron from the primary vertex (PV) with $p_T$ greater than 28 GeV, within $|\eta| < $ 2.4. The isolation of each electron is computed by adding the $p_T$ of all tracks within a cone of size $\Delta R < 0.3$, not including the $p_T$ of the electron, divided by the sum of the $p_T$ of all tracks within the same $\Delta R$ cone. Here, $\Delta R$ is computed as $\sqrt{\Delta \eta^2 + \Delta \phi^2}$, where $\Delta \phi$ and $\Delta \eta$ are the differences in azimuthal angle and pseudorapidity, respectively, between the electron and the tracks. For the current study, a fairly isolated electron is required, with an isolation factor (sum of $p_T$ of tracks divided by sum of $p_T$ of all tracks) less than 0.1. The trigger thresholds for our study are adopted from the L1 trigger menu designed for the HL-LHC, as outlined in reference \cite{CERN-LHCC-2020-004}.

    \item \textbf{Single TkIsoMuon}- Requires at least one prompt, isolated muon with $p_T >$ 22 GeV from PV, within $|\eta| < $ 2.4. The isolation of each muon is calculated in the same way as explained above for the electron trigger, i.e., by summing the $p_T$ of all tracks within a $\Delta R < 0.3$ cone around the muon, excluding the muon's own $p_T$, divided by the sum of the $p_T$ of all tracks within the same $\Delta R$ cone. Similarly, for this trigger, the muon isolation factor is required to be less than 0.1. The trigger thresholds used in our study are obtained from the L1 trigger menu for the HL-LHC as provided in reference \cite{CERN-LHCC-2020-004}.
    \end{itemize}

Various thresholds for object $p_T$, isolation, $H_T$ and jet timing for above mentioned L1 triggers are summarised in Table \ref{tab:L1triggers}.

\begin{figure}[hbt!]
    \centering
    \includegraphics[width=\textwidth]{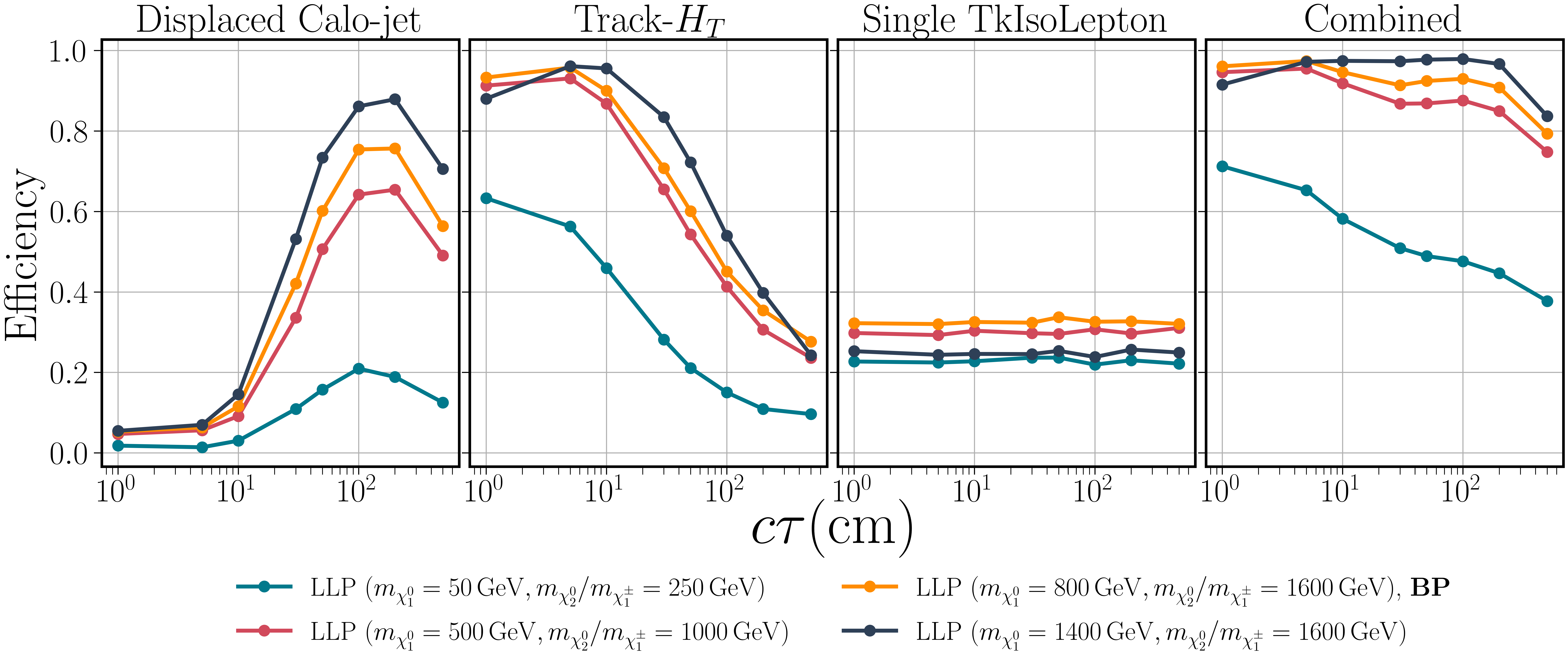}~~
    \caption{Variation of trigger efficiency for displaced Calo-Jet, Track-$H_T$, and single TkIsoLepton triggers with decay length for four LLP scenarios with one benchmark scenario (BP). The combined trigger efficiency is also shown.}
    \label{fig:trig_eff}
\end{figure}

\begin{table}[t]
\centering
\begin{tabular}{|c|c||}
\hline
L1 Trigger & Online thresholds \\ 
\hline\hline
IsoElectron & $p_T >$ 28 GeV, $Iso <$ 0.1 \\ 
\cline{1-2}
IsoMuon & $p_T >$ 22 GeV, $Iso <$ 0.1  \\ 
\cline{1-2}
Track $H_T$ & $H_T >$ 160 GeV, $p_T^{jet} >$  5 GeV   \\ 
\cline{1-2}
Calo-jet & $\Delta T >$ 1.2 ns, $p_T^{jet} >$  35 GeV, $N_{tow} \geq$ 4\\
\hline
\hline
\end{tabular} 
\caption{Selection cuts for the L1 triggers}
\label{tab:L1triggers}
\end{table}
Figure \ref{fig:trig_eff} displays the variation of trigger efficiency with mean proper decay length for the triggers mentioned above, as well as the combined trigger efficiency for four different LLP scenarios. Although, we only consider LLPs with masses higher than 500 GeV in the current analysis, trigger efficiency for LLPs with masses ranging from light ($M_{\chi_1^{0}}$ = 50 GeV) to very heavy($M_{\chi_1^{0}}$ = 1400 GeV) is shown to depict the variation of trigger efficiency with mass of LLP. We also show the variation of trigger efficiency for one of the benchmark point with $M_{\chi_2^{0}}/M_{\chi_1^{\pm}}$ = 1600 GeV and $M_{\chi_1^{0}}$ = 800 GeV. LLPs with decay lengths ranging from 1 cm to 500 cm and originating from the decay of $\chi_2^0/\chi_1^{\pm}$ with masses varying from 250 GeV to 1600 GeV are considered.
 The observations obtained from Figure \ref{fig:trig_eff} are summarized as follows:
\begin{itemize}

\item The displaced Calo-Jet trigger is particularly effective for LLPs with longer decay lengths, especially for heavier LLPs. This is because the trigger utilizes timing information from the ECAL deposits, and LLPs that decay later in the detector will exhibit a more significant time delay. As the mass of the LLPs increases, the time delay also increases, as these heavier LLPs travel at a slower pace, resulting in a more considerable time delay.

\item Furthermore, the track $H_T$ trigger performs best for LLPs with smaller decay lengths, as the efficiency of extended track reconstruction degrades with increasing c$\tau$. The trigger efficiency is also reduced for LLP benchmarks with smaller mass differences between $\chi_2^0/\chi^{\pm}$ and their decay products due to less hadronic activity in the calorimeter.

\item The lepton trigger efficiency remains unaffected by the c$\tau$ parameter, but it decreases as the mass degeneracy between the LLP and $\chi_2^0/\chi^{\pm}$ increases, due to kinematic suppression. This implies that the efficiency of lepton triggers will be lower for benchmark points with smaller mass differences between the LLP and $\chi_2^0/\chi^{\pm}$.

\item The combined trigger efficiency decreases with increasing c$\tau$ for every benchmark point, which can be explained by looking at the individual trigger efficiencies. The decrease in efficiency as the decay length increases is likely because LLP events with longer decay lengths have a higher chance of escaping the detection region before being triggered.

\item Single TkIsoLepton triggers, along with the Track-$H_T$ trigger and displaced Calo-Jet trigger, complement each other in selecting LLP events in both the lower and higher ends of the decay length spectrum for both lighter and heavier LLPs. This implies that combining these triggers can effectively select LLP events across a wide range of decay lengths and masses.

\end{itemize}

It is important to highlight the significance of displaced jet triggers in detecting LLP events, especially since lepton triggers are limited to selecting events with prompt leptons. In the current study, prompt leptons mainly come from the inclusive decay of W and Higgs bosons, which have relatively low branching fractions. For instance, in a scenario where the LLP originates from the decay of a 1000 GeV particle with a mass of 500 GeV and a decay length of 10 cm, the efficiency of the lepton trigger is approximately 30\%. However, the overall efficiency increases significantly when displaced jet triggers are included. With a decay length of 10 cm, the efficiency rises to around 91\%, and for a longer decay length of 100 cm, the efficiency remains high at 89\%. This demonstrates that incorporating displaced jet triggers significantly enhances the efficiency of detecting LLPs with shorter as well as longer decay lengths. In LLP scenarios with shorter decay lengths, the track $H_T$ trigger is more effective, while in contrast, the Calo-Jet trigger is more effective for events with longer decay lengths. In conclusion, the most effective approach to efficiently select LLP events with varying decay lengths, from very small to very large, is to use a combination of different L1 triggers.

\section{Offline analysis}
\label{sec:offana}

After triggering the events at L1, the next step is to analyze the selected events offline to remove the background events that have huge cross-sections. We begin by reconstructing the secondary displaced vertex, a key characteristic of the decay of LLPs, for the selected events using the set of displaced tracks. In our analysis, we reconstruct tracks taking into consideration the track reconstruction efficiency, which varies with the transverse displacement of tracks from the beam-line as achievable at Phase-I of the LHC for CMS \cite{disptrackHLT} since no specific information about the track reconstruction efficiency is available for Phase-II. However, we assume that offline track reconstruction in terms of transverse displacement from the beam-line will remain the same for Phase-II as in Phase-I. Nevertheless, updated information will be needed to confirm this assumption. We form displaced vertices by clustering displaced tracks with transverse impact parameter $|d_0|>$ 1.5 mm based on their spatial position. We identify vertices with at least two displaced tracks associated with the vertex. Each vertex is assigned a unique ID and stored for further analysis.
Next, we compute two physics variables related to the each selected displaced vertex- 
\begin{itemize}
    \item $\mathbf{N_{trk}^{disp}}$ - The number of displaced tracks associated with the secondary vertex.
    \item $\mathbf{M_{DV}}$ - The invariant mass of the displaced secondary vertex, which is calculated using the displaced tracks that are associated with it.
\end{itemize}

\begin{figure}[hbt!]
    \centering

    \includegraphics[scale=0.36]{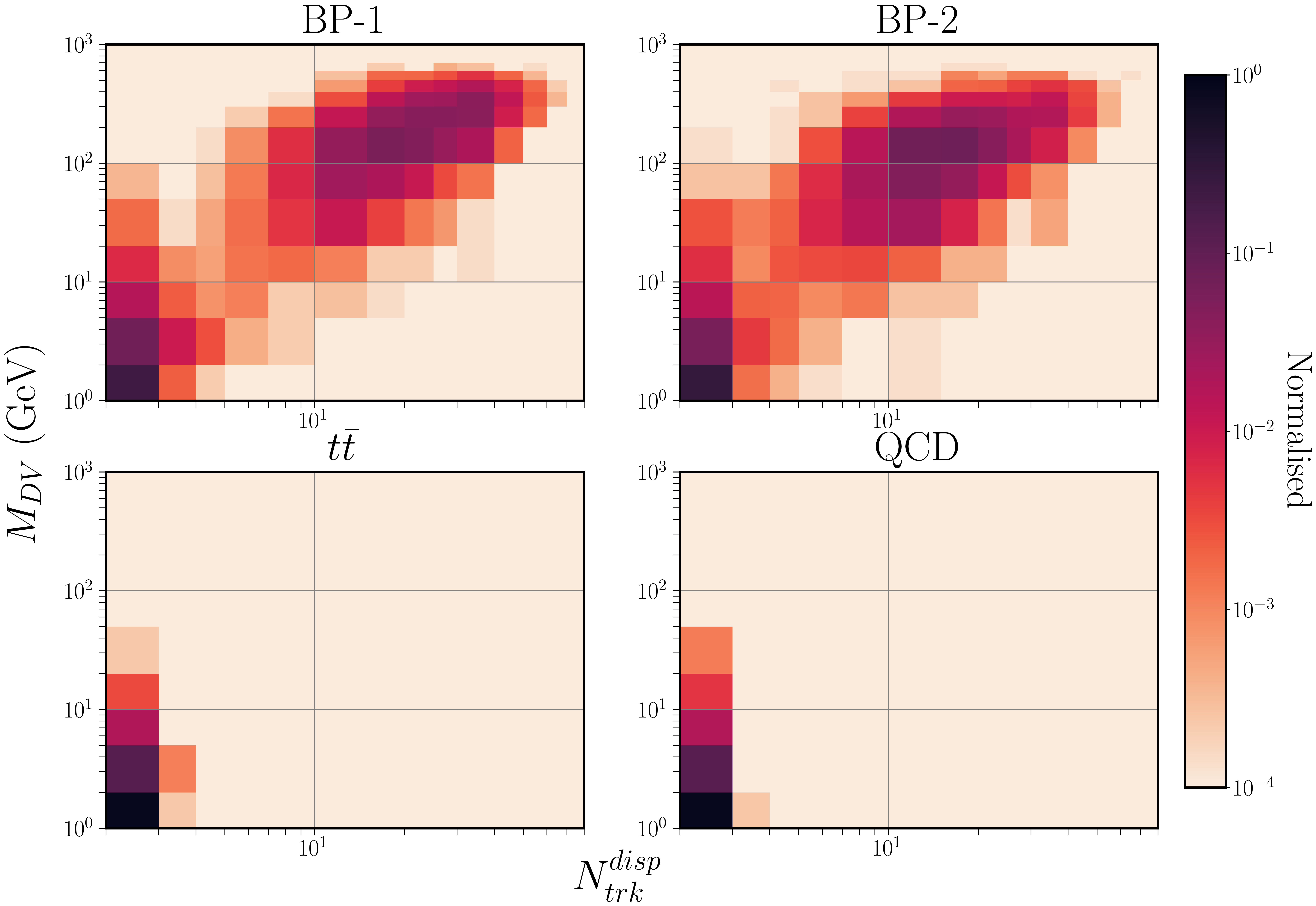}
    \caption{Two-dimensional distribution showing the relationship between the number of displaced tracks ($N_{trk}^{disp}$) and the displaced vertex invariant mass ($M_{DV}$) for the two LLP benchmarks, BP-1 and BP-2, along with the $t\bar{t}$ and QCD background.}
    \label{fig:disp_track1}
\end{figure}

In Figure \ref{fig:disp_track1}, we show the two-dimensional distribution of displaced track multiplicity ($N_{trk}^{disp}$) and invariant mass of the displaced vertex ($M_{DV}$). The distributions are shown for two LLP benchmark points, BP-1 and BP-2, as well as for the QCD and $t\bar{t}$ background. To ensure proper normalization of the data, each bin in the distribution is re-weighted such that the sum of the fraction of entries falling in every bin equals unity.

As shown in Figure \ref{fig:disp_track1}, the LLP benchmarks exhibit a significantly higher number of displaced tracks associated with the displaced vertex as the invariant mass of the displaced vertex increases compared to the backgrounds. This indicates that applying a suitable 2-dimensional cut on the displaced track multiplicity and the invariant mass of the displaced vertex can effectively reduce the contribution from the background events. 
In addition to mitigating the background events from QCD and $t\bar{t}$, implementing a higher threshold cut on both $N_{trk}^{disp}$ and $M_{DV}$ can effectively remove the displaced vertices originating from the instrumental background as shown in \cite{ATLAS:2023oti,CMS-PAS-EXO-19-013,CMS:2020iwv}. This is because the displaced vertices from the instrumental background are typically collimated and have lower multiplicity and smaller invariant mass than the signal. As shown in \cite{ATLAS:2023oti}, instrumental background can be effectively mitigated by the requirement on $M_{DV}$ and $N_{trk}^{disp}$ where they implement a threshold of 10 GeV on $M_{DV}$ and a threshold of 5 for $N_{trk}^{disp}$ in the signal region. For the signal, the invariant mass of the displaced vertex is expected to peak around the mass of the LLP, which in this case is 800 GeV. However, it is essential to note that the number of reconstructed displaced tracks may be reduced for very short ($\approx 1 \,\mathrm{cm}$) or very long decay lengths ($\approx 500\, \mathrm{cm}$), which can impact signal efficiency. 

Now, we turn our attention to the utilisation of MTD timing information in the current analysis. At the HL-LHC, MTD will be positioned between the tracker and the electromagnetic calorimeter of the CMS detector, providing precise timing information for the charged particles originating within the tracker. Currently, precision timing information from the MTD is proposed to be included for the offline analysis in the CMS detector instead of the online trigger system. Including the partial readout of the MTD for a region of interest at L1 is a possibility in the future upgrades of the HL-LHC \cite{CMS:2667167}. However, in this work, we have mainly focused on including output from MTD in the offline analysis, where we can construct complex physics variables out of the output from various sub-detectors, including MTD. At HL-LHC, the primary objective of the MTD will be to help mitigate the effect of the huge amount of PU on physics analysis and restore the physics performance at par with Phase-I of LHC. However, the role of MTD will be pivotal in studying exotic particles such as LLPs, where the decay of the particles is delayed, and timing information from the MTD can be efficiently used to search for such particles. 

Timing information can be extracted from the MTD with the timing resolution of 30 ps for MTD hits from the charged particles with $p_T >$ 0.7 GeV in the barrel region ($|\eta| <$ 1.5) and $p >$ 0.7 GeV in the endcap region (1.5 $ < |\eta| < $ 3.0). Excellent coverage and timing resolution of the MTD can be leveraged to construct the timing variables for the jets originating due to the decay of LLPs which will be delayed in time. MTD layer is proposed to be placed at the radius of 1.16 m between the tracker and barrel ECAL, which is placed at the radius of 1.29 m.   

In order to construct timing variables for jets using information from MTD, we will require MTD hits directly below the clustered jets within the specific cone along the jet axis. In addition to MTD hits coming from tracks, whether displaced or prompt, we have two additional lists of MTD hits - one where MTD hits originate from reconstructed displaced tracks with $|d_0|>$ 1.5 mm and the second one where we have MTD hits with no reconstructed tracks associated with the hits. We construct physics variables using the three abovementioned collections of MTD hits. We consider MTD hits only within a narrow cone radius directly below the jets to reduce the PU contamination. For MTD hits associated with tracks, we only consider tracks with $p_T>$ 2 GeV and can be reconstructed using the track construction efficiency as explained before. We have constructed following timing variables using above mentioned three MTD hits collection directly below a clustered jet in a cone with $\Delta R <$ 0.3 and axis matching with the jet axis-

\begin{itemize}
    \item $\mathbf{N_{MTD}}$: The number of MTD hits with associated reconstructed tracks within R = 0.3 of the jet axis. Hard signal jets will contain comparatively higher number of MTD hits when compared to the background. LLPs decaying after the MTD and before ECAL and HCAL boundary will have energy deposition in the calorimeters but with no associated MTD hits. So, MTD hit multiplicity will decrease with the increase in the decay length; however, distribution will mainly be dominated by the charged PU hits. 

    \item $\mathbf{N_{MTD}^{Disp}}$: The number of MTD hits with associated reconstructed tracks within R = 0.3 of the jet axis with $|d_0|>$ 1.5 mm. Displaced jets will contain comparatively higher number of MTD hits coming from displaced tracks when compared to the background. MTD hit multiplicity will decrease with the increase in the decay length.
    
    \item $\mathbf{N_{MTD}^{NT}}$: The number of MTD hits within R = 0.3 of the jet axis with no associated tracks. Track reconstruction efficiency follows an efficiency curve where the efficiency of reconstructing a track will degrade with the transverse distance ($D_{xy}$) from the beam line. As a result, we will have a higher number of MTD hits with no associated tracks for displaced LLPs. However, this number will decrease with the decreasing decay length as we will have more and more tracks with smaller $D_{xy}$ being reconstructed. In contrast, for prompt processes, most of the MTD hits will have associated tracks; hence $N_{MTD}^{NT}$ will be less than displaced LLPs.

    \item $\mathbf{T_{Raw}}$: The mean of the timing of MTD hits constituting a jet within cone radius of $R = 0.3 $. To compute $T_{raw}$, no timing calibration corresponding to the position of MTD hits has been applied. For highly displaced LLPs, $T_{raw}$ will have higher values compared to prompt processes, but since the majority of MTD hits inside a jet will be coming from PU interactions, $T_{raw}$ measurement will mainly be dominated by the timing of PU hits. Also, the timing of the jet will depend on the position and $p_T$ of the jets. Jets with low $p_T$ depositing energy at higher $\eta$ values away from the central part of the barrel will have higher timing which is valid for both LLPs and the prompt background processes. 

     \item $\mathbf{T_{Raw}^{Disp}}$: The mean of the timing of MTD hits associated with displaced tracks constituting a jet within cone radius of $R = 0.3 $. For highly displaced LLPs, $T_{raw}$ will have higher values compared to prompt processes where displaced track multiplicity is very low.
    
    \item $\mathbf{T_{Raw}^{NT}}$: The mean of the timing of MTD hits with no associated tracks within cone radius of $R = 0.3$ of the jet axis. As we discussed earlier, most tracks can be successfully reconstructed for prompt processes and LLPs with very short decay lengths; therefore, jets from such processes are less likely to leave MTD hits with no reconstructed tracks. In such cases, $T_{raw}^{NT}$ will be zero when no MTD hit is found with no reconstructed tracks. However, with the increase in decay length, we will have more and more number of MTD hits with no reconstructed tracks. Furthermore, for prompt processes, contribution to the tail of the timing distribution of the jet will be coming from the hits with very low $p_T$ tracks, which did not get reconstructed. 
    
    \item $\mathbf{T_{Calib}}$: The mean of the timing of the MTD hits within R = 0.3 of the jet axis calibrated with respect to origin (0,0,0). Calibration of the temporal position of each hit is done to mitigate the effect of the position of the MTD hit in the $\eta-\phi$ plane on the timing of the MTD hit. The timing of each MTD hit is corrected such that if the particle travels with the speed of light from the origin to the position of the MTD hit, it should take zero seconds to reach there. Hence, the timing of the delayed particles will be given as the difference between the raw timing of the hit, as discussed before, and the time taken by a massless particle travelling with the speed of light originating from the origin to reach the position of the MTD hit. $T_{calib}$ will have higher values and longer tail in the timing distribution for displaced jets than those coming from prompt processes.
    
    \item $\mathbf{T_{Calib}^{Disp}}$: The mean of the timing of the MTD hits associated with displaced tracks as explained above within R = 0.3 of the jet axis calibrated with respect to origin (0,0,0). $T_{calib}$ will have higher values and longer tail in the timing distribution for displaced jets compared to jets coming from prompt processes where displaced track multiplicity will be very low compared to displaced processes.
    
    \item $\mathbf{T_{Calib}^{NT}}$: The mean of the calibrated timing of the MTD hits with no associated tracks within R = 0.3 of the jet axis. As explained earlier, prompt processes and displaced particles with very small decay lengths will have MTD hits which can be easily associated with the reconstructed tracks. More and more number of MTD hits will be available for highly displaced particles with no reconstructed tracks to be fed into the calculation of $T_{calib}$. As a result, the timing distribution of the displaced jets will have slightly higher values of $T_{calib}$ when compared to prompt processes. However, $T_{calib}$ will have smaller values as we consider LLPs with shorter and shorter decay lengths. For prompt processes, contribution to the tail of the timing distribution of the jet will be coming from the hits with very low $p_T$ tracks which did not get reconstructed. 
    
    \item $\mathbf{p_T^{Ratio}}$: The Ratio of the sum of $p_T$ of reconstructed tracks (prompt as well as displaced) associated with MTD hits within R = 0.3 of the jet and the corresponding jet $p_T$. For LLPs with large decay lengths, fewer and fewer prompt tracks will be reconstructed, and hence the number of MTD hits with no associated tracks will be smaller. As a result, there will be a more significant mismatch between actual jet $p_T$ and $p_T$ calculated using tracks with associated hits. This effect will be minimal for prompt processes where most MTD hits will have associated tracks.

    \item $\mathbf{D_{T}^{Med}}$: The median of the transverse distance calculated using the reconstructed tracks that have hits in the MTD and are associated with the jets within $\Delta R<$ 0.3.
\end{itemize}

\begin{figure}[hbt!]
    \centering
    \includegraphics[scale=0.35]{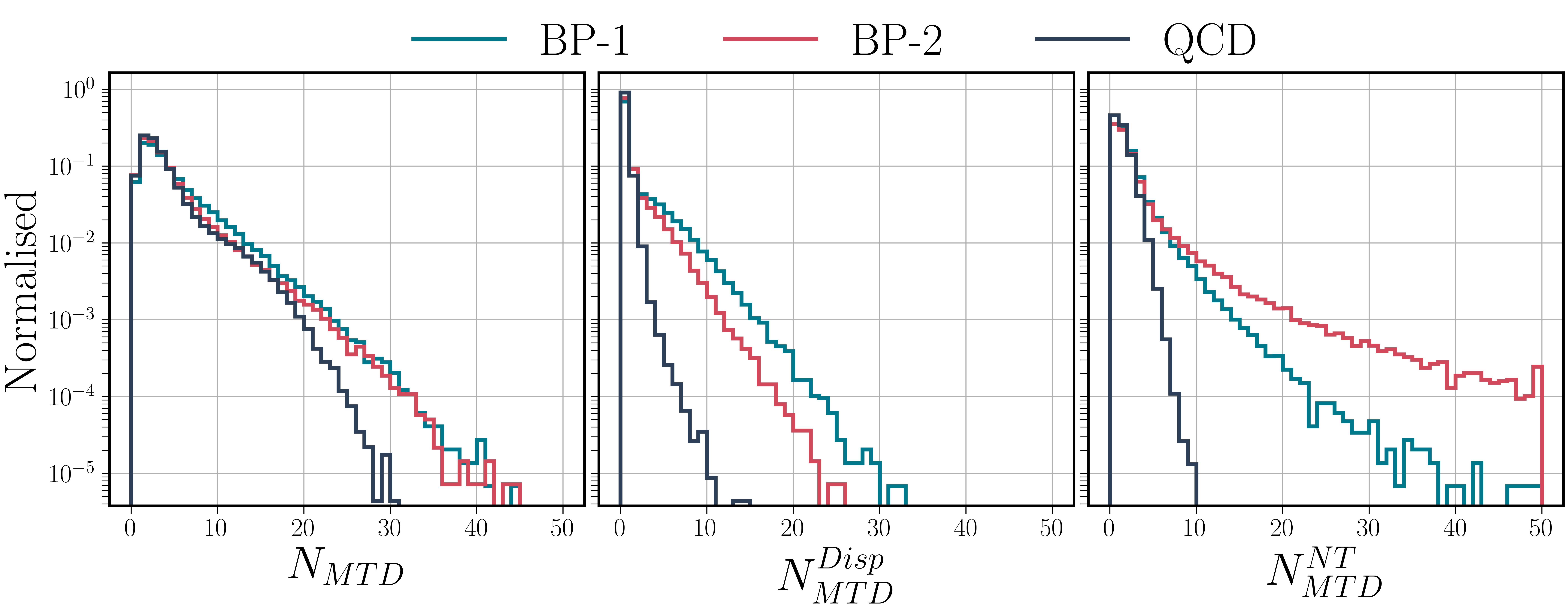}~~
    \caption{Distribution of the Multiplicity of MTD ($N_{MTD}$) hits for three MTD hits collections, for the QCD background, and the two LLP benchmark points BP-1 and BP-2, at the HL-LHC.}
    \label{fig:nmtd_mtd}
\end{figure}

\begin{figure}[hbt!]
    \centering
    \includegraphics[scale=0.35]{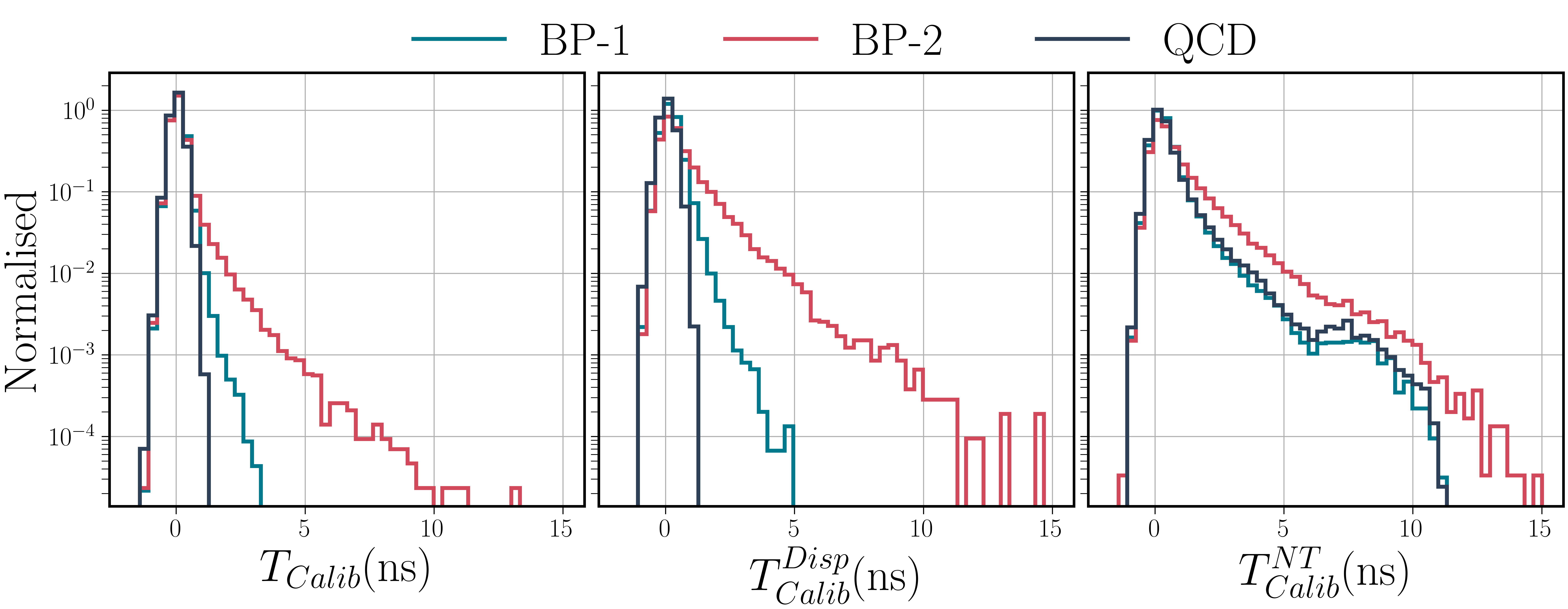}~~
    \caption{Calibrated time ($T_{Calib}$) calculated using three MTD hits collections, for the QCD background, and the two LLP benchmark points BP-1 and BP-2, at the HL-LHC.}
    \label{fig:tcalib_mtd}
\end{figure}

\begin{figure}[hbt!]
    \centering
    \includegraphics[scale=0.35]{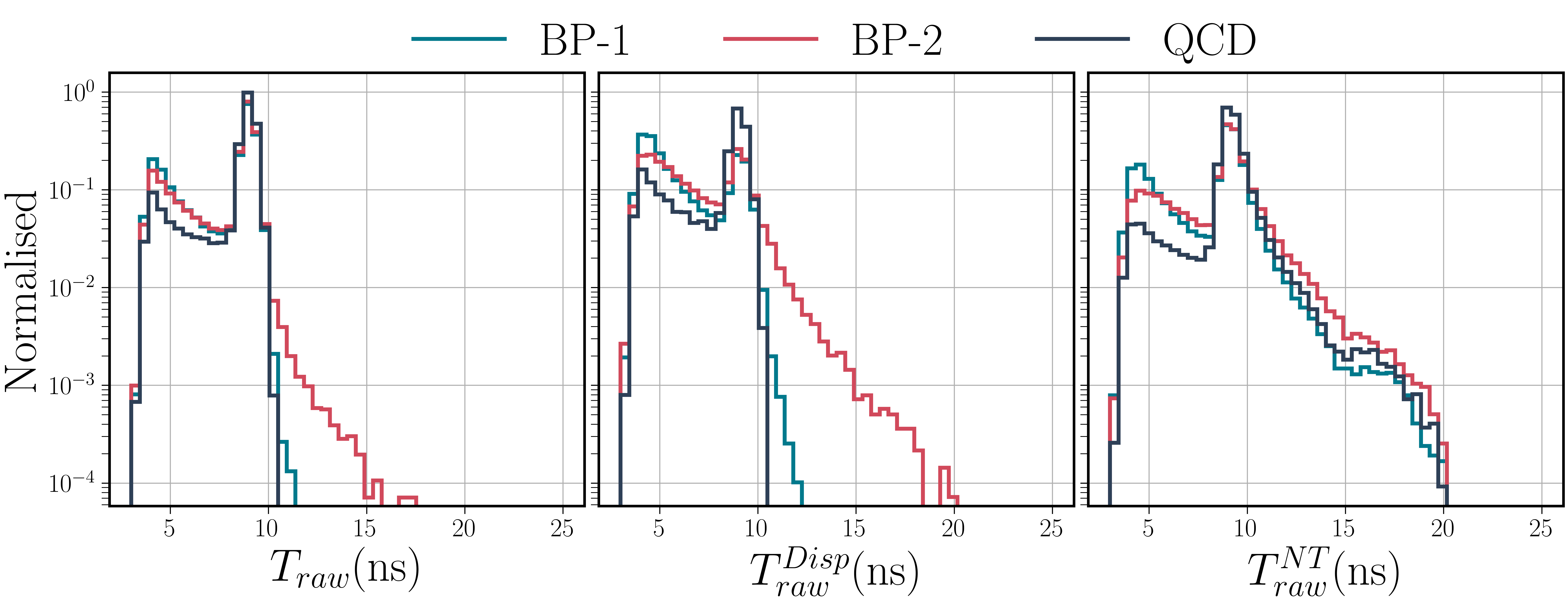}~~
    \caption{Raw time ($T_{Raw}$) calculated using three MTD hits collections, for the QCD background, and the two LLP benchmark points BP-1 and BP-2, at the HL-LHC.}
    \label{fig:traw_mtd}
\end{figure}

Figure \ref{fig:nmtd_mtd} shows the multiplicity of MTD hits for each jet, as measured using three lists of MTD hits for LLP benchmarks BP-1 and BP-2 and the QCD background under the conditions of HL-LHC. Similarly, Figure \ref{fig:tcalib_mtd} and Figure \ref{fig:traw_mtd} depict the $T_{Calib}$ and $T_{Raw}$, respectively, calculated using the three MTD hits collections.

\begin{figure}[hbt!]
    \centering
    \includegraphics[scale=0.4]{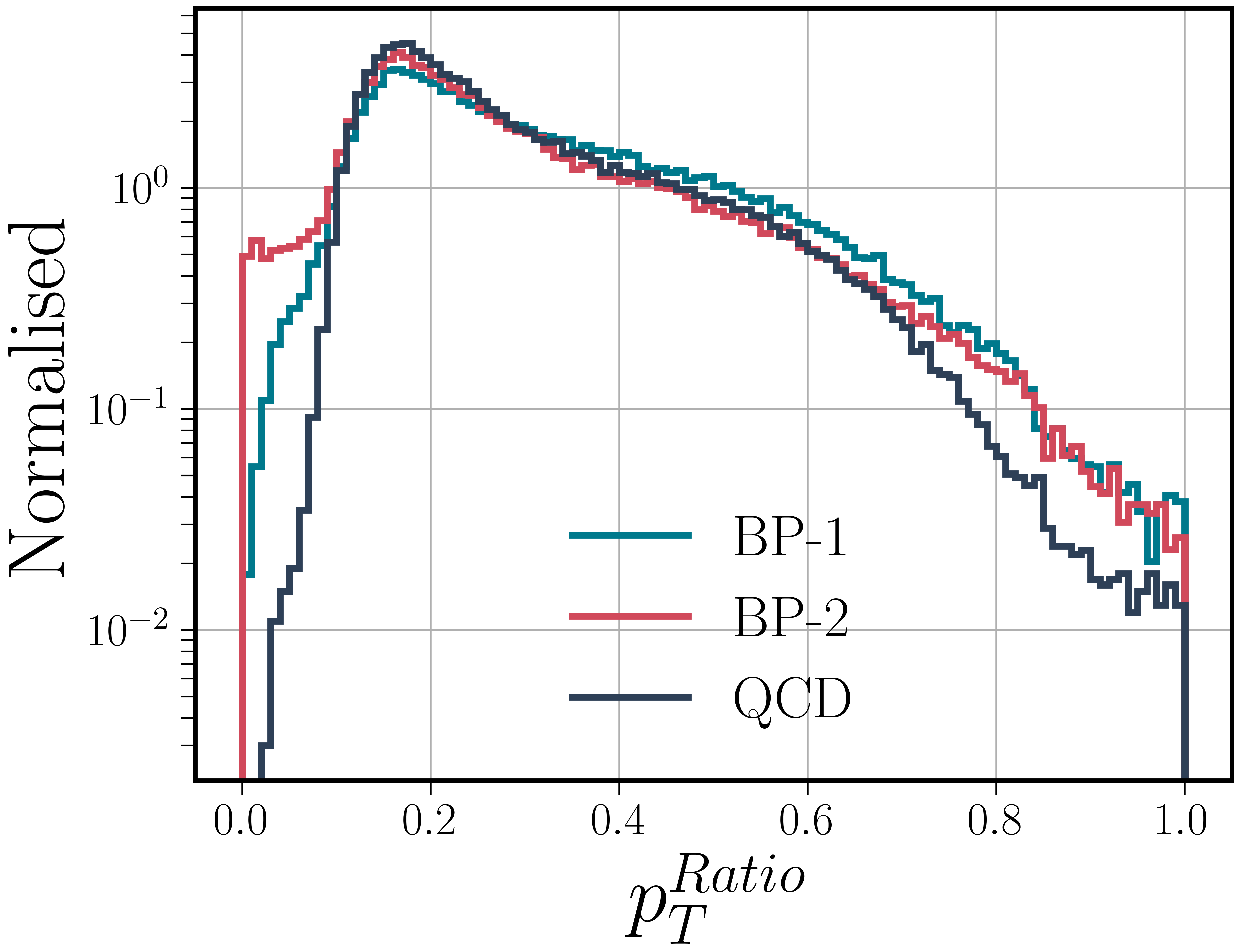}~~
    \caption{Ratio of sum of $p_T$ of tracks associated with MTD hits within $\Delta R <$ 0.3 of the calorimeter jet and jet $p_T$, as calculated using calorimeter inputs using the anti-$k_T$ jet algorithm with R=0.3, for the QCD background, and the two LLP benchmark points BP-1 and BP-2, at the HL-LHC.}
    \label{fig:ptraio_mtd}
\end{figure}

From Figure \ref{fig:nmtd_mtd},  \ref{fig:tcalib_mtd} and  \ref{fig:traw_mtd} We observe-
\begin{itemize}
    \item LLP with smaller decay length has more number of MTD hits compared to LLP with higher decay length. LLPs in general has more number of MTD hits compared to background sources. 
    \item Number of MTD hits with associated displaced tracks decrease with increasing LLP decay length while for background, number is significantly lower due to absence of displaced tracks inside the jet.
    \item LLP with higher decay length have more number of hits with no associated tracks compared to LLP with lower decay length while backgrounds have very less number of MTD hits with associated tracks since most of the hits in MTD will be coming from promptly produced particles.
    \item Tail of the timing distribution ($T_{Calib}$ and $T_{Raw}$) calculated using MTD hits with associated tracks and MTD hits associated with displaced tracks only increases with decay length for LLPs and background and signal can be easily distinguished.    
    \item Timing calculated using MTD hits with no associated tracks has longer tail for LLP with higher decay lengths. High timing in timing distribution calculated using MTD hits with no associated tracks in background is associated with low $p_T$ PU tracks which move very slowly and contaminate the jet timing.  
    \item Tail at the lower end of the $p_T^{ratio}$ is observed for LLPs because of the mismatch between jet $p_T$ and $p_T$ calculated using tracks associated with MTD hits. Effect is more pronounced for higher decay length because of the higher probability of not finding hits with associated tracks. 
\end{itemize}

Now, we will shift our attention to the timing of ECAL tower constituents within jets to construct various timing variables for jets. For jet formation, we require ECAL and HCAL towers with energy deposits $E_{em}>$ 0.5 GeV and $E_{had}>$ 1 GeV, respectively. Timing is calculated only for those jets with at least one ECAL tower exceeding an energy deposit of 1 GeV. For timing calculation, we only take into account ECAL towers by requiring $E_{had}<$ 0.0001 GeV and $E_{em}>$ 0.5 GeV. In Section \ref{sec:trigger}, we have already utilized one of the timing variables, namely the energy-weighted mean timing of the jet ($\Delta T_{mean}^{Ewt}$), which is used in the design of the L1 trigger based on ECAL timing. Additionally, we have computed several other measures for the jets using the ECAL timing. These measures are listed as follows: 

\begin{itemize}
\item $\mathbf{\Delta T_{mean}}$: The average timing of all ECAL crystals associated with the jet as shown in Equation \ref{eq:delTmean}. Here, $i$ refers to all ECAL crystals within the jet, and $N$ is the complete count of the crystals associated with the jet.
\begin{equation}
\Delta T_{mean} = \frac{\sum \Delta T_{i}}{N},
\label{eq:delTmean}
\end{equation}
\item $\mathbf{\Delta T_{median}}$: The median timing of all ECAL crystals associated with the jet.

\item $\mathbf{\Delta T_{RMS}}$: The root mean square value of the timing of all ECAL crystals with in the jet as computed in Equation \ref{eq:trms}.
\begin{equation}
\Delta T_{RMS} = \sqrt{\frac{\sum \Delta T_{i}^2}{N}},
\label{eq:trms}
\end{equation}

\item $\mathbf{\sum \Delta T}$: The sum of the timing of all ECAL crystals in the jet.

\item $\mathbf{\Delta T_{mean}^{Ewt}}$: The energy-weighted mean timing of all ECAL crystals in the jet. This is computed as the sum of the product of each crystal's timing and energy divided by the total energy of all crystals within the jet, as in Equation \ref{eq:delTmean-ewt}. 
\begin{equation}
\Delta T_{mean}^{Ewt} = \frac{\sum \Delta T_{i}\times E_{i}}{\sum E_{i}}
\label{eq:delTmean-ewt}
\end{equation}

\item $\mathbf{\Delta T_{mean}^{ETwt}}$: The transverse energy-weighted mean timing of all ECAL crystals in the jet as shown in Equation \ref{eq:delTmean-eTwt}. 
\begin{equation}
\Delta T_{mean}^{ETwt} = \frac{\sum \Delta T_{i}\times E_{T,i}}{\sum E_{T,i}}
\label{eq:delTmean-eTwt}
\end{equation}

\end{itemize}

Before re-weighting with energy or transverse energy for the aforementioned timing variables, we adjust the timing of each ECAL crystal relative to the origin, as explained in \ref{sec:trigger}.

We have also implemented two more different calibration techniques for the above-mentioned timing variables where we calibrate the timing of each crystal in the jet with respect to the primary vertex (PV) and the jet vertex (JV). The PV is determined by using prompt track collection. The vertex with the highest $\sum p_T^2$ is selected as the PV. Similarly, the JV is determined by considering all prompt tracks associated with the jet, located within a distance of $\Delta R<0.3$ from the jet axis at the ECAL. The vertex with the maximum $\sum p_T^2$ is chosen as the JV.

Additionally, the mean timing of a jet is computed using only five or ten crystals, with the maximum time delay determined by multiplying the maximum value of time delay with the energy of the crystal, denoted as ($\Delta T\times E)_{mean}^{Max 5}, (\Delta T\times E)_{mean}^{Max 10}$. The mean timing of the jet calculated using 5 and 10 most energetic crystals is denoted as $\Delta T_{mean}^{Max 5}$ and $\Delta T_{mean}^{Max 10}$, respectively. We additionally compute two quantities, namely $(\Delta T \times E)_{mean}^{TMax 5}$ and $(\Delta T \times E)_{mean}^{EMax 5}$ where we calculate mean timing of a jet using only five ECAL towers, using the maximum value of time delay multiplied by the energy of the crystals, and this product is divided by the timing and energy of the five ECAL towers possessing highest energy and timing values respectively.

For quantities calculated above, If the jet contains less than five or ten towers, the values of $\Delta T_{mean}^{Max 5}$ and $\Delta T_{mean}^{Max 10}$ are assigned the same values as $\Delta T_{mean}$, while the values of $(\Delta T\times E)_{mean}^{Max 5}$ and $(\Delta T\times E)_{mean}^{Max 10}$ as well as $(\Delta T \times E)_{mean}^{TMax 5}$ and $(\Delta T \times E)_{mean}^{EMax 5}$ are assigned the same values as $\Delta T_{mean}^{Ewt}$. Introducing such variables in the analysis is crucial as they are more resistant to PU contamination. Additionally, using crystals with the highest $\Delta T\times E$ values make sure that PU hits with low energy and high ECAL timing do not significantly affect these variables.

We also compute several other quantities using information about the tracks and calorimeter towers associated with the jet-  

\begin{itemize}
\item $\mathbf{p_T^{Ratio}}$- Sum of the $p_T$ of all tracks associated with the jet within a distance of $\Delta R < 0.3$ from the jet axis, divided by the jet $p_T$ as determined through calorimeter inputs using anti-$k_T$ jet alogrithm with R=0.3.
\item $\mathbf{\Delta \eta, \Delta \phi}$ - Differences in the ($\eta,\phi$) position of a jet as calculated using tracks and ECAL towers within the jet. ($\eta, \phi$) of the jet using tracks is calculated by using $p_T$ weighted mean of ($\eta,\phi$) of tracks contained within $\Delta R  <$ 0.3 of the jet. Similarly,  ($\eta, \phi$) of the jet is calculated using position of ECAL crystals contained with in the jet by re-weighting them with crystal $E_T$. For displaced LLPs, position of jets constructed using available tracks will differ from the position of the jets constructed using ECAL crystals since less and less number of displaced tracks will be reconstructed with the increase in the decay length of the LLPs.
\item $\mathbf{\sum E_{tow}}$ - Sum of energy of ECAL towers with in $\Delta R = $ 0.3
\item $\mathbf{\frac{E_{had}}{E_{jet}}}$ - Fraction of energy deposited in hadron calorimeter (HCAL) compared to total jet energy.
\item $\mathbf{N_{trk, prompt}^{jet}}$ - Number of prompt tracks associated with jet located within a $\Delta R$ of less than 0.3 from the jet axis.
\item $\mathbf{N_{trk, disp}^{jet}}$ - Number of displaced tracks with $|d_0|>$1.5 mm associated with jet within $\Delta R <$ 0.3 of jet axis.
\end{itemize}

\begin{figure}[hbt!]
    \centering
    \includegraphics[scale=0.35]{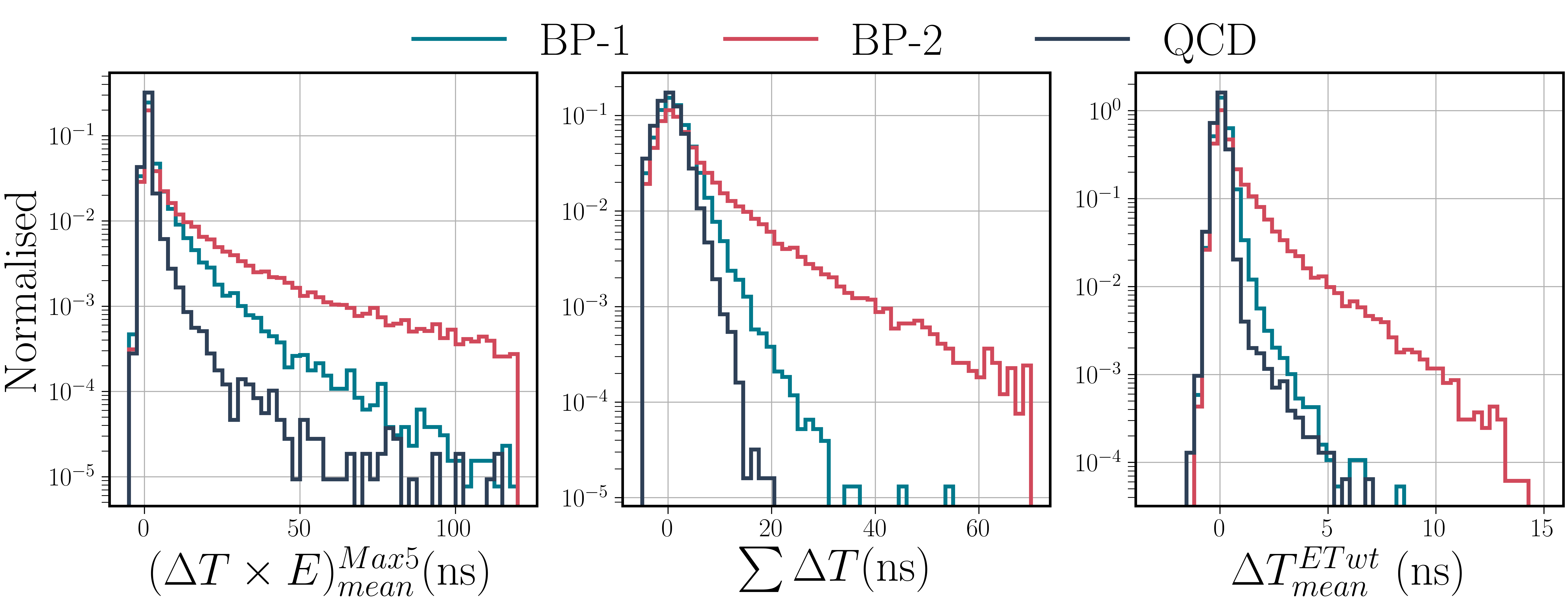}~~
    \caption{Energy-weighted mean timing of a jet, calculated exclusively from the 5 crystals having the maximum time delay, $(\Delta T \times E)_{mean}^{Max 5}$ (left), sum of the timing all ECAL crystals associated with a jet, $\sum \Delta T$ (middle), and the transverse energy-weighted mean timing of the jet $\Delta T_{mean}^{ETwt}$ (right) for QCD background, and the two LLP benchmark points BP-1 and BP-2, at the HL-LHC. }
    \label{fig:ecal_var}
\end{figure}

In Figure \ref{fig:ecal_var}, we show the distributions of three important timing variables constructed using the information from ECAL, namely $(\Delta T\times E)_{mean}^{Max 5}$, $\sum \Delta T$, and $\Delta T_{mean}^{ETwt}$, for QCD background and two LLP benchmark points, BP-1 and BP-2.

As we can see from the Figure \ref{fig:ecal_var}, LLP benchmarks exhibit a longer tail in the timing distribution when compared to the QCD background as expected. Compared to BP-1, where the LLP has a decay length of 10 cm, discrimination is more pronounced for BP-2, where the LLP has a decay length of 100 cm. The plot for $\sum \Delta T$, which shows the sum of the time delay for all the hits in the ECAL, also demonstrates a significant difference between the LLP benchmarks and the background, with a more prominent difference for BP-2. Timing variable $(\Delta T \times E)_{mean}^{Max 5}$ and $\sum \Delta T$ show comparatively better discrimination between jet timing for QCD and LLP when LLP with lower decay length is considered.

Now, we study the correlation between different timing variables constructed using ECAL timing. Our aim is to identify variables that exhibit high correlation factors for both signal and background, while contributing little to distinguish between them. Such redundant variables can be omitted from the analysis thus improving efficiency and interpretability of the analysis. Figure \ref{fig:corr_llp} and \ref{fig:corr_qcd} in Appendix \ref{app:corr} illustrates the correlation matrix for the LLP benchmark BP-2  and the QCD background with $H_T^{gen} = \{500-600\}$ GeV respectively. As we can see from the figure, several variables show strong correlation with each other for both signal and background. Such variables can be termed redundant and thus excluded from the final analysis. On the other hand, there are some variables showing a strong correlation for signal while exhibiting weak correlation for background; such variables can be helpful in distinguishing the signal from the background.

Now, with the definitions of various physics variables as described above, we divide the offline analysis of the events selected through the triggers defined in Section \ref{sec:trigger} into three separate and independent parts, as described below:
\begin{itemize}
    \item \textbf{Cut-based analysis (CBA)}: In order to get a handle on LLP scenarios with shorter decay lengths, we adopt a cut-based approach to efficiently select signal events while significantly rejecting background contribution. We apply an appropriate two-dimensional cut on two variables: $N_{trk}^{disp}$, which represents the number of displaced tracks associated with the secondary vertex, and $M_{DV}$, which represents the invariant mass of the displaced vertex as defined earlier. 
    \item \textbf{Multi-variate analysis-1 (MVA-1)}: To get a handle on the events with significant lifetime, a machine learning-based multi-variate analysis denoted as MVA-1 is performed independently on the jets from the events selected at L1 using variables constructed using the information from the MTD and associated information from the tracker as calculated previously. The variables used in this analysis are tabulated in Table \ref{tab:variables}.
   \item \textbf{Multi-variate analysis-2 (MVA-2)}: Similar to MVA-1, we conduct a separate MVA analysis referred to as MVA-2 on the jets from the events selected at L1 aimed at LLP scenarios with large lifetime where we utilize variables that are constructed using information from the ECAL as well as associated data from the tracker, as calculated previously. The variables used for this analysis are listed in Table \ref{tab:variables}.
   
\end{itemize}

\begin{table}[hbt!]
\centering
\resizebox{\columnwidth}{!}{%
\begin{tabular}{|c|c|c|c||}
\hline
\textbf{Analysis} &
  \textbf{CBA} &
  \textbf{MVA-1} &
  \textbf{MVA-2} \\ \hline \hline
\textbf{Variables} &
  $H_T$, $N_{trk}^{disp}$, $M_{DV}$ &
  \begin{tabular}[c]{@{}c@{}}$N_{MTD}$, $N_{MTD}^{Disp}$, $N_{MTD}^{NT}$, \\ $T_{raw}$, $T_{raw}^{Disp}$, $T_{raw}^{NT}$,\\ $T_{calib}$, $T_{calib}^{Disp}$, $T_{calib}^{NT}$,\\ $p_T^{Ratio}$, $p_T^{jet}$, $\eta^{jet}$\end{tabular} &
  \begin{tabular}[c]{@{}c@{}}$\Delta T_{mean}$, $\Delta T_{mean}^{PV}$, $\Delta T_{mean}^{PVJ}$, $\sum \Delta T$,\\  $\sum \Delta T^{PV}$, $\sum \Delta T^{PVJ}$, $\Delta T_{mean}^{ewt}$, $\Delta T_{mean}^{ewt,PV}$,\\ $\Delta T_{mean}^{ewt,PVJ}$, $\Delta T_{mean}^{etwt}$, $\Delta T_{mean}^{etwt,PV}$, $\Delta T_{mean}^{etwt,PVJ}$,\\ $\Delta T_{median}$, $\Delta T_{median}^{PV}$, $\Delta T_{median}^{PVJ}$, $\Delta T_{RMS}$,\\ $\Delta T_{RMS}^{PV}$, $\Delta T_{RMS}^{PVJ}$, $\Delta T^{Max5}_{mean}$, $(\Delta T \times E)^{Max5}_{mean}$,\\ $(\Delta T \times E)^{TMax5}_{mean}$, $(\Delta T \times E)^{EMax5}_{mean}$,\\ $\Delta T^{Max10}_{mean}$, $(\Delta T \times E)^{Max10}_{mean}$,\\ $p_T^{Ratio}$, $\frac{E_{had}}{E_{jet}}$, $\sum E_{tow}$, $N_{trk, prompt}^{jet}$, \\ $N_{trk, disp}^{jet}$,  $p_T^{jet}$, $\eta^{jet}$ \end{tabular} \\ \hline \hline
\end{tabular}%
}
\caption{Different physics variables to be used in cut-based analysis (CBA) and two independent multi-variate analyses (MVA-1 and MVA-2) }
\label{tab:variables}
\end{table}

Dividing the analysis into three independent parts with CBA focused on LLPs with smaller decay lengths and MVA-1 and MVA-2 focused on LLPs with larger decay lengths helps us address the challenge associated with background suppression and signal extraction across a wide range of decay lengths, from very small to very large. Utilizing these three approaches ensures the analysis remains sensitive to various LLP scenarios with a spectrum of decay lengths.

The final signal significance for each LLP benchmark point is calculated by combining results from the abovementioned approaches after removing the duplicate events. In the following sections, we will provide a detailed explanation of the analysis approaches mentioned above.

\subsection{Cut-based analysis (CBA)}
\label{subsec:cba}
Owing to the inclusion of events from displaced Calo-Jet and track-$H_T$ trigger at L1, the significant contribution to the background will originate from the jets coming from instrumental effects, the QCD processes and $t\bar{t}$ events. Contribution from these background sources can be significantly reduced by requiring events with high $H_T$ where we calculate $H_T$ by summing over the $p_T$ of all the jets in each event. Furthermore, as previously stated, an appropriate two-dimensional threshold cut on $M_{DV}$ and $N_{trk}^{disp}$ will also lead to a significant reduction in the background events. For the cut-based analysis, we apply the following selection cuts:
\begin{itemize}
    \item \textbf{Event} $\mathbf{H_T}$: We require events selected at L1 to possess event $H_T$ greater than 500 GeV where $H_T$ is calculated using jets with jet $p_T>$ 40 GeV. 
    \item $\mathbf{N_{trk}^{Disp}}$: We require at-least one reconstructed secondary vertex with at-least six associated displaced tracks, each with a transverse impact parameter ($|d_0|$) greater than 1.5 mm. 
    \item $\mathbf{M_{DV}}$: We require the invariant mass of the reconstructed secondary vertex to be greater than 20 GeV.
\end{itemize}

Events selected after imposing the abovementioned cuts are sorted and stored for further analysis, where we combine results from MVA-1 and MVA-2 with CBA. Now, let us discuss the second and third approaches, MVA-1 and MVA-2.

\subsection{Multivariate analysis -1 (MVA-1)}
\label{subsec:mva-1}
In MVA-1, we utilise an XGBoost (Extreme Gradient Boosting) \cite{Chen:2016:XST:2939672.2939785} model trained on physics variables constructed using MTD information to specifically target LLPs with longer lifetimes. XGBoost works by iteratively building a series of decision trees, where each tree corrects the errors made by the previous trees. To minimize the loss function, which measures the difference between the predicted and actual values, XGBoost uses a gradient descent optimization algorithm. In our analysis, we use the following set of XGBoost parameters to train our model for a multi-class classification :
\begin{itemize}
    \item \textbf{objective}: The objective of our model was to perform multi-class classification using the `multi:softprob' approach, which computes the predicted probabilities for each class.
    \item \textbf{num\_class}: This parameter was set to 8, indicating the total number of classes in our multi-class classification problem with class 0 representing the signal while other 7 classes ranging from 1 to 5 representing QCD background in different $H_T^{gen}$ bins while class 6 and 7 representing $t\bar{t}$ and W+Jets background respectively.
    \item \textbf{eval\_metric}: We used `mlogloss' as our evaluation metric, which calculates the multi-class logarithmic loss during the training process. It provides a measure of the model's performance.
    \item \textbf{learning\_rate}: We utilized a learning rate of 0.1, which determines the step size at each boosting iteration.
    \item \textbf{early\_stopping\_rounds}: We implemented early stopping with a value of 5 for this parameter. This means that if the loss does not decrease further after 5 consecutive iterations, the training process is halted. The purpose of early stopping is to prevent over-training and improve the generalization ability of the model. 
    \item \textbf{colsample\_bytree}: This parameter was set to 0.3, indicating the fraction of columns to be randomly sampled for each tree during training. 
    \item \textbf{max\_depth}: We set the maximum depth of each tree in our model to 6. This restricts the depth of the individual trees, preventing overfitting and improving generalization.
    \item \textbf{alpha}: The alpha parameter was assigned a value of 4, which controls the L1 regularization term on the weights. It helps in reducing the complexity of the model and preventing overfitting.
    \item \textbf{tree\_method:} The `tree\_method' parameter was set to 'gpu\_hist', indicating the use of GPU acceleration for training the model. 
    \item \textbf{num\_boost\_rounds}: To ensure convergence of the training and achieve the minimum loss, we set the number of boosting rounds to 1000 epochs. This value determined the maximum number of iterations performed during the training process.

\end{itemize}

The XGBoost model is trained using a set of variables described in the third column of Table \ref{tab:variables}. We focus on LLP benchmark points with $M_{\chi_2^0}/M_{\chi_1{\pm}}$=1600 GeV, and vary the mass of the LLP, $M_{\chi_1^0}$, from 500 GeV to 1000 GeV.

Events selected at L1 utilising displaced and lepton triggers are further required to have offline event $H_T$ greater than 500 GeV where event $H_T$ is calculated in similar manner as for CBA as explained in Section \ref{subsec:cba}. Only 6 leading jets per event with transverse momentum greater than 100 GeV are considered in the training process to exclude most pileup jets and only keep the jets coming from the hard interaction. The decay length of the LLPs in our benchmark scenarios ranges from 1 cm to 500 cm. To account for different decay lengths, we train three separate XGBoost models, each targeting a specific range of decay lengths. 

Three models are trained using LLP benchmark scenarios exhibiting decay length of 1 cm, 50 cm and 200 cm to target LLPs with decay length in range of 1 cm to 5 cm, 10 cm to 50 cm and 100 cm to 500 cm respectively. We choose the mass of the LLP, $M_{\chi_1^0}=$ 800 GeV for training the XGBoost models for each decay length, as it falls within the moderate range of LLP masses considered in our analysis, with  $M_{\chi_2^0}/M_{\chi_1{\pm}}$ fixed at 1600 GeV. Each jet in the training sample for the background is assigned a weight according to process cross-section and number of generated events for that particular sample such that sum of weights is unity. Signal jets are assigned unit weights. The trained XGBoost models are then utilized to classify LLPs in the respective decay length ranges in the subsequent analysis steps. 

We divide the jets selected at L1 and after the pre-selection cuts, as defined above, into training and testing datasets of equal size. We have approximately 3600k jets from $t\bar{t}$ events and 2200k jets from QCD dijet events, the two dominant background sources. For signal benchmark points with decay lengths of 1 cm, 50 cm, and 200 cm, we have approximately 5800k, 5400k, and 4600k jets respectively.

\begin{figure}[hbt!]
    \centering
    \includegraphics[scale=0.4]{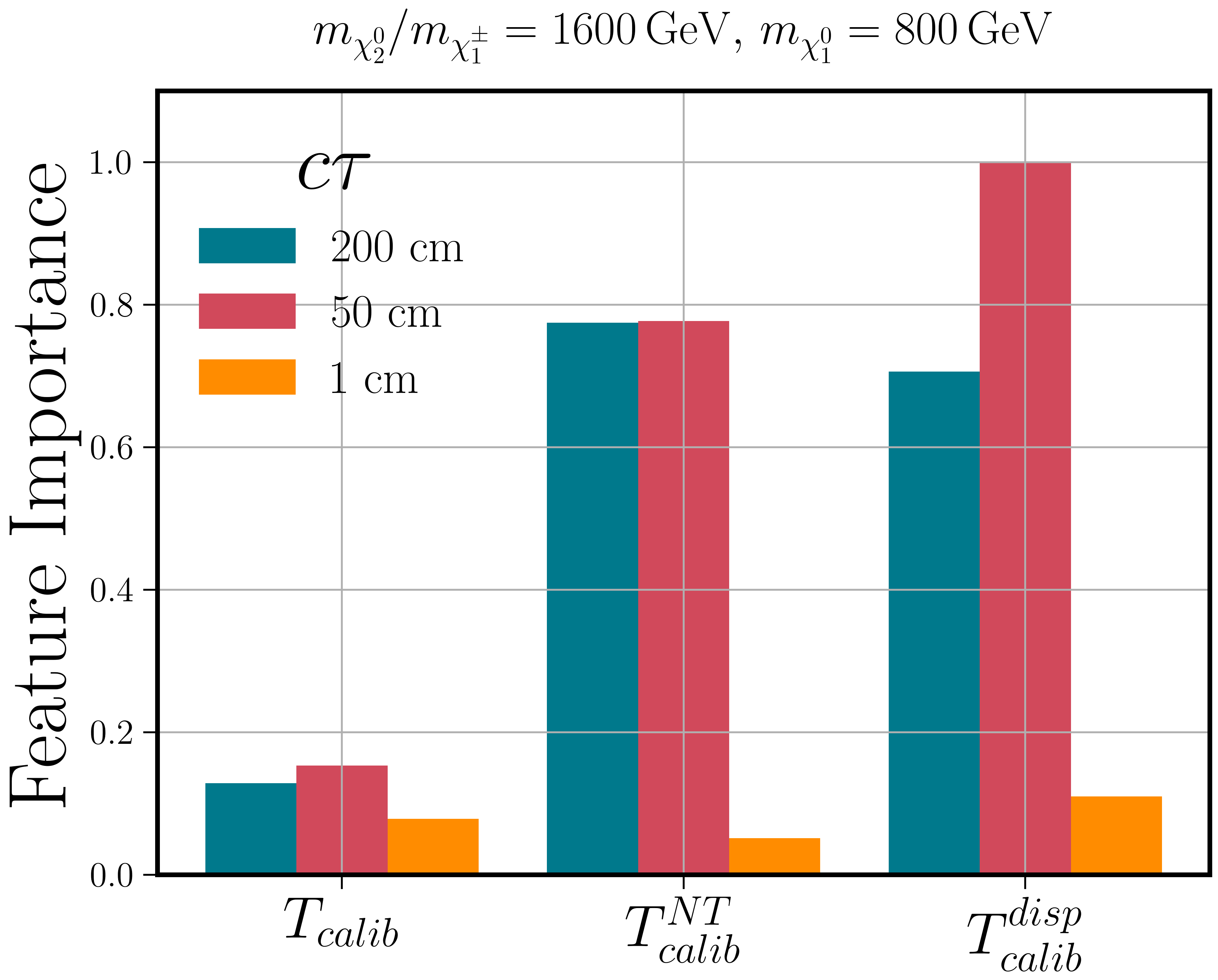}~~
    \includegraphics[scale=0.38]{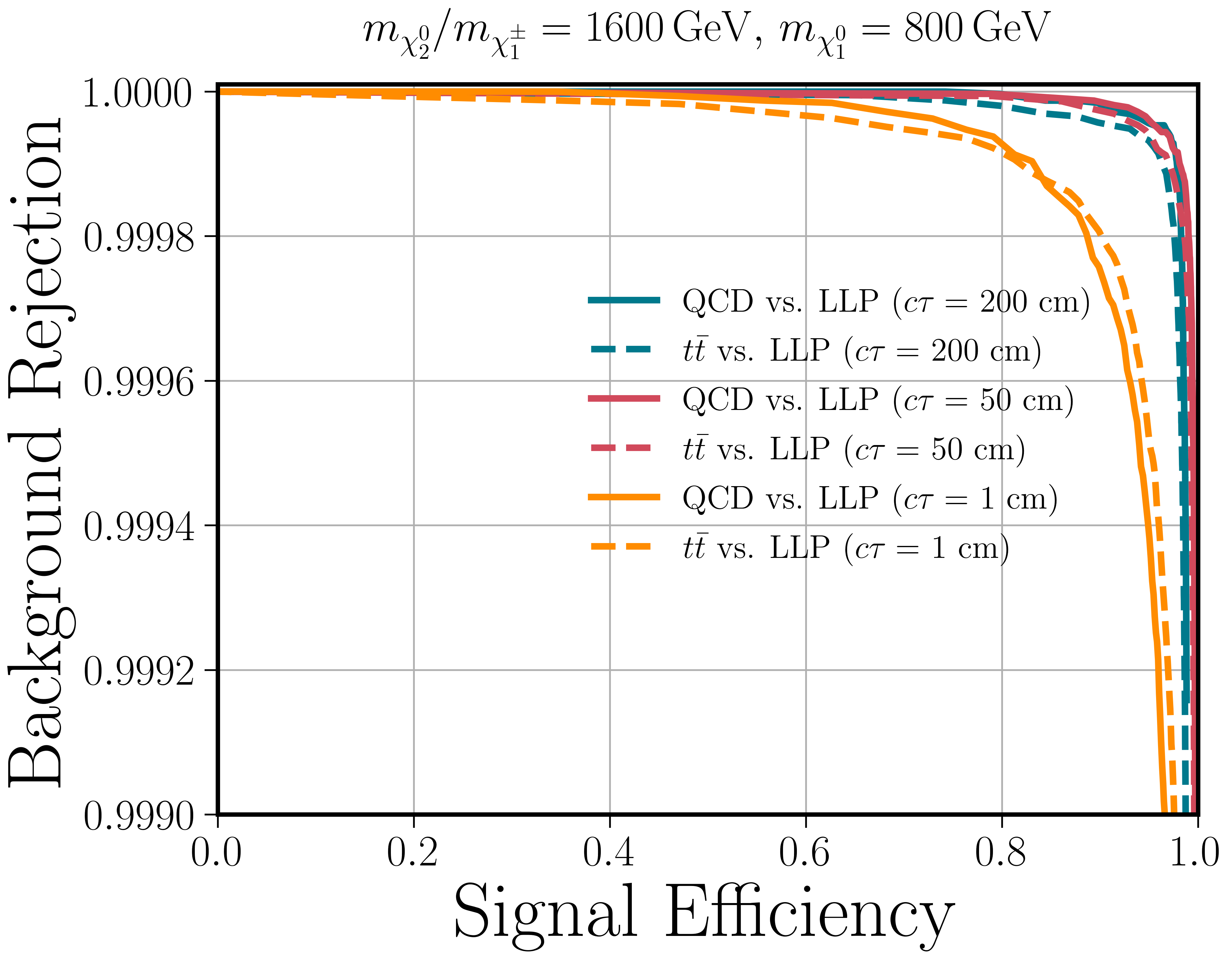}
    \caption{Relative feature importance of three important variables of MVA-1 for three LLP scenarios where decay length is 1 cm, 50 cm and 200 cm (\textit{left}) and classification in terms ROC for two dominant background (QCD and $t\bar{t}$) for LLP with decay length 1 cm, 50 cm and 200 cm (\textit{right}) .}
    \label{fig:feature_roc_mtd}
\end{figure}

To highlight the significance of timing information in selecting LLP events with large lifetime, we show the feature importance of three crucial variables for the MTD in three different LLP scenarios, with decay lengths of 1 cm, 50 cm, and 200 cm in Figure \ref{fig:feature_roc_mtd} (\textit{left})..
Feature importance is evaluated using the gain metric, which quantifies the improvement in accuracy achieved by a feature in the decision tree branches. In the case of MTD, timing information is derived from tracks (displaced or prompt) that leave hits in the MTD or from hits with no associated tracks. As shown in Figure  \ref{fig:feature_roc_mtd} (\textit{left}), Jet timing calculated using MTD hits with associated tracks performs well for LLP scenarios where the decay length allows for the reconstruction of a larger number of tracks, including displaced ones. However, for LLPs with very long decay lengths, the timing of jets calculated using MTD hits with no associated tracks gains more significance due to the abundance of MTD hits without associated tracks.  

In Figure \ref{fig:feature_roc_mtd} (\textit{right}), we present the signal efficiency versus background rejection in terms of Receiver Operating Characteristic (ROC) curves for three different decay lengths, namely 1 cm, 50 cm, and 200 cm, for LLPs with $M_{\chi_2^0}=1600$ GeV and $M_{\chi_1^0}=$ 800 GeV. The plots demonstrate that the MVA-1 approach, which incorporates timing information from the MTD, exhibits significantly improved performance for LLP scenarios with longer decay lengths compared to those with shorter decay lengths while maintaining good performance for LLP scenarios with shorter decay lengths. This improvement can be attributed to the inclusion of timing information from MTD, which aids in better discriminating between signal and background events, particularly for LLPs with longer decay lengths. 

To finalize the event selection, we impose the prerequisite of at least one jet in every event that exhibits a very high signal probability. This signal probability is determined based on the amount of background rejection required, which will depend on the decay length of the LLP on which the model was trained.

\subsection{Multivariate analysis -2 (MVA-2)}
\label{subsec:mva-2}
For MVA-2, we follow the same training strategy as outlined in the previous section for MVA-1. However, we utilize a different set of variables to train the XGBoost models, listed in the fourth column of Table \ref{tab:variables}. In this case, physics variables are constructed using timing information from the ECAL instead of MTD, as was done for MVA-1. Similar to MVA-1, the main objective of MVA-2 is to identify LLPs with longer lifetimes effectively.

For MVA-2, we have approximately 2200k jets from $t\bar{t}$ events and 1400k jets from QCD dijet events, which are the two dominant background sources. As for the signal benchmark points with decay lengths of 1 cm, 50 cm, and 200 cm, we have approximately 5600k, 5200k, and 4400k jets respectively.

\begin{figure}[hbt!]
    \centering
    \includegraphics[scale=0.4]{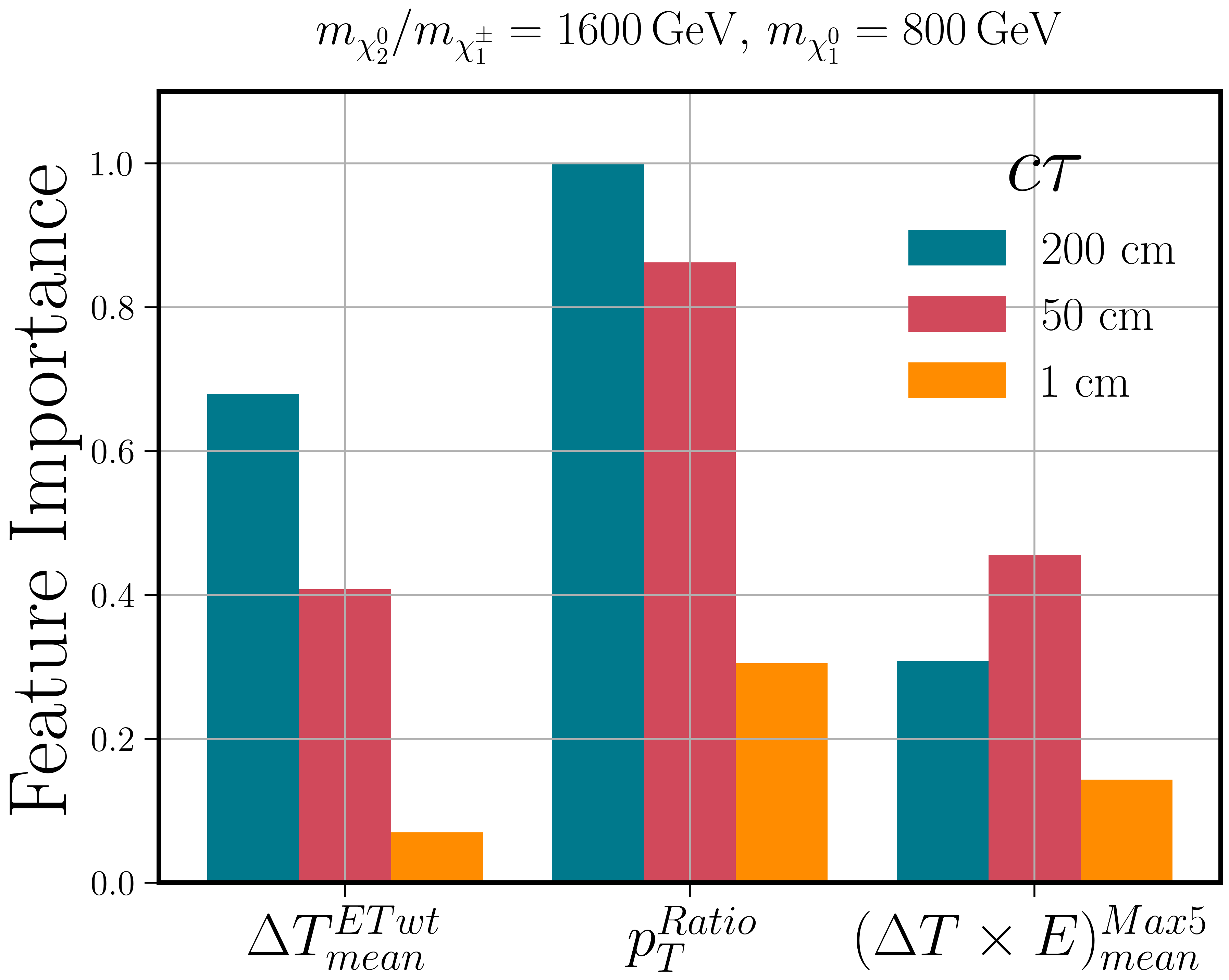}~~
    \includegraphics[scale=0.38]{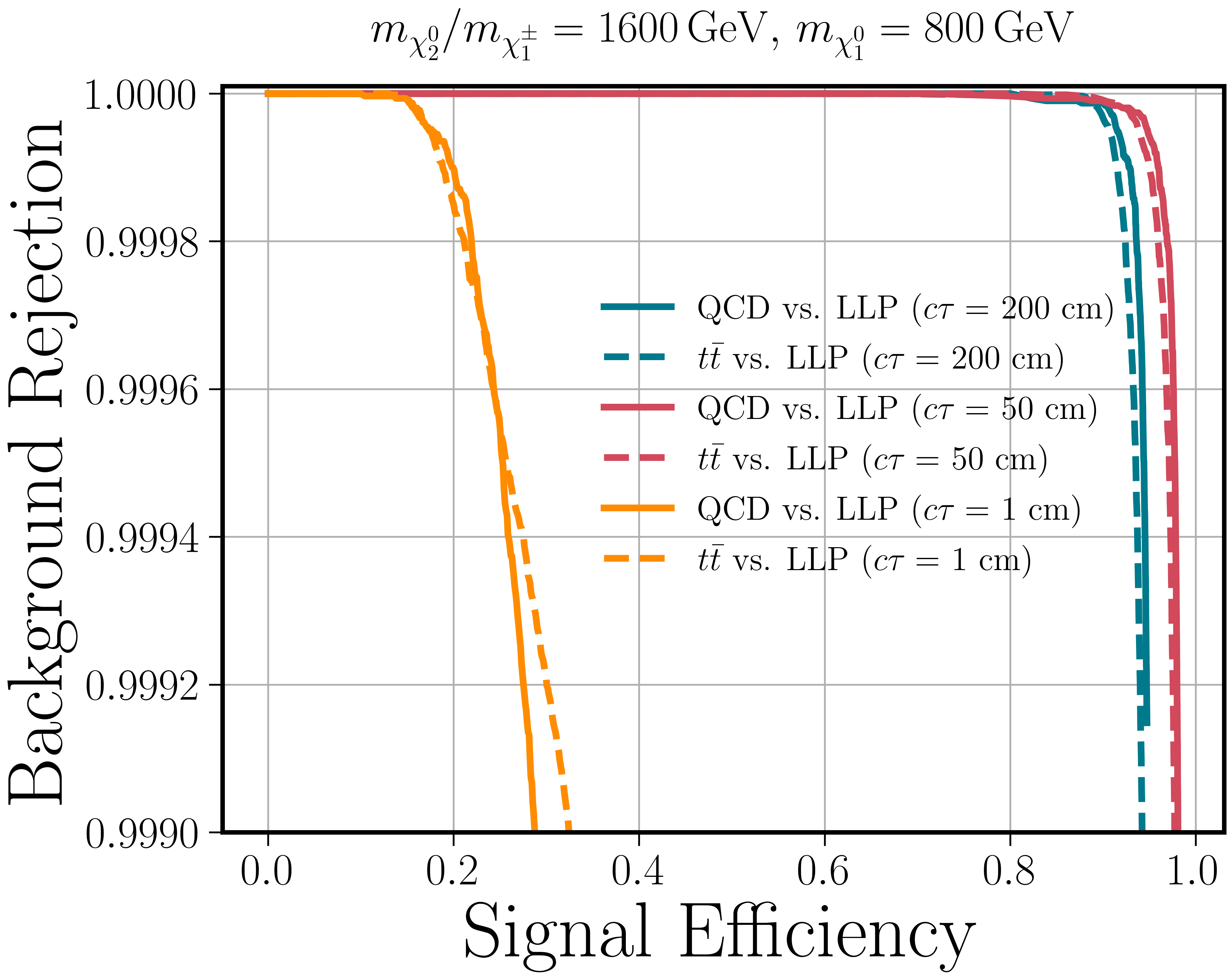}
    \caption{Relative feature importance of three important variables of MVA-2 for three LLP scenarios where decay length is 1 cm, 50 cm and 200 cm (\textit{left}) and classification in terms ROC for two dominant background (QCD and $t\bar{t}$) for LLP with decay length 1 cm, 50 cm and 200 cm (\textit{right}) .}
    \label{fig:feature_roc_ecal}
\end{figure}

We now study the performance of three crucial physics variables included in MVA-2 regarding their relative importance in classifying jets in signal and background for three LLP scenarios with decay lengths of 1 cm, 50 cm and 200 cm. Similar to MVA-1, we utilize the gain metric to quantify the importance. The relative feature importance of these variables is shown in Figure \ref{fig:feature_roc_ecal} (left).

As we can see from Figure \ref{fig:feature_roc_ecal} (\textit{left}), An higher relative importance is assigned to the timing variables, $\Delta T_{mean}^{etwt}$ and $(\Delta T \times E)^{Max5}_{mean}$, for LLPs with larger decay lengths compared to LLPs with smaller decay lengths, as expected. Similarly, $p_T^{Ratio}$ holds more significance for LLPs with larger decay lengths than LLPs with smaller decay lengths. This discrepancy can be understood from the fact that a more significant mismatch arises between the jet $p_T$ calculated using tracks within the jet and the calorimeter jet $p_T$ as the LLP decay length increases, resulting from the fewer displaced tracks being reconstructed. 

Here, we would also like to highlight the importance of incorporating energy or transverse energy re-weighting when calculating the timing of the jet. Energy-weighted timing variables exhibit higher significance in classification than timing variables without energy re-weighting. This difference arises from the fact that considering energy-weighted quantities helps mitigate the PU contamination in the jet timing. Since PU energy deposits are soft, their effect on the timing of the jet is reduced after taking their energy into account to construct the jet timing.

In Figure \ref{fig:feature_roc_ecal} (\textit{right}), we show the ROC curves for three different decay lengths of LLP considering QCD and $t\bar{t}$ background separately, considering the same LLP benchmark scenario described in the previous section. We can observe that MVA-2 outperforms the LLPs with decay lengths of c$\tau =$ 50 cm and 200 cm, compared to 1 cm, emphasizing the vital role of ECAL timing in distinguishing highly displaced LLPs from the background. 

In order to make a final selection of events, we impose a criterion that requires at least one jet in each event to have a very high signal probability. This effectively eliminates the majority of the jets originating from background sources.

Next, we will quantify the results obtained from CBA, MVA-1 and MVA-2 in terms of signal significance.

\section{Results}
\label{sec:results}
The final signal significance is determined by combining the outcomes of MVA-1, MVA-2, and CBA, while ensuring that duplicate events are excluded from the final event selection. The signal (S) or background (B) yield is calculated using the following equation:

\begin{equation}
S \text{ or } B = \sigma_{\text{process}} \times \epsilon \times \mathcal{L}
\end{equation}

where $\sigma_{\text{process}}$ represents the production cross-section of the process, $\epsilon$ represents the selection efficiency, and $\mathcal{L}$ represents the integrated luminosity. The selection efficiency is determined by dividing the number of finally selected events, obtained after combining the results from MVA-1, MVA-2, and CBA, by the total number of events. In this analysis, integrated luminosity of 3000 $\mathrm{fb^{-1}}$ for HL-LHC is considered.

Finally, for each signal benchmark point, we calculate signal significance using the following formula:
\begin{equation}
S_{\text{sig}} = \frac{S}{\sqrt{B}}
\end{equation}
where $S_{\text{sig}}$ represents the signal significance, and $S$ and $B$ represent the signal and background yields, respectively. 

In Table \ref{tab:cutflow}, we present the number of events, yield, and signal significance obtained from CBA, MVA-1 and MVA-2 for three LLP benchmark points with decay lengths of 1 cm, 50 cm, and 200 cm, and $M_{\chi_2^0}=1600$ GeV and $M_{\chi_1^0}=$ 800 GeV, along with three background sources.

\begin{table}[hbt!]
\small
\centering
\begin{tabular}{|cc|l|c|c|c|c|c|c|}
\hline
\multicolumn{2}{|c|}{Events}                                                                                                                       & \begin{tabular}[c]{@{}l@{}}Total \\ (mil)\end{tabular} & CBA     & MVA-1     & MVA-2      & Combined & Yield & $S_{sig}$ \\ \hline
\multicolumn{1}{|c|}{\multirow{3}{*}{\begin{tabular}[c]{@{}c@{}}LLP\\ $M_{\chi_1^0} = 800$ GeV\\ $M_{\chi_2^0} = 1600$ GeV\end{tabular}}} & 1cm    & 0.5                                                     & 444681 & 29154 & 49406   & 447699 &  215 &     9.28                                              \\ \cline{2-9} 
\multicolumn{1}{|c|}{}                                                                                                                    & 50 cm  & 0.5                                                      & 404420 & 330181 & 365242  & 455612  &  219 &     9.46                                                   \\ \cline{2-9} 
\multicolumn{1}{|c|}{}                                                                                                                    & 200 cm & 0.5                                                      & 219169 & 274411 & 333449  & 415166  &  200 &     8.59                                                   \\ \hline
\multicolumn{2}{|c|}{QCD}                                                                                                                          & 3                                                        & 0      & 0      & 0      & 0    &  0 &     -                                                       \\ \hline
\multicolumn{2}{|c|}{$t\bar{t}$}                                                                                                                   & 5                                                      & 0      & 0      & 1      & 1    &  534.0 &     -                                                       \\ \hline
\multicolumn{2}{|c|}{W+Jets}                                                                                                                       & 1                                                        & 0      & 0      & 0      & 0   &  0 &     -                                                       \\ \hline
\hline
\end{tabular}
\caption{Total number of events for signal and background obtained individually from the CBA, MVA-1, and MVA-2 analyses, as well as the combined number of events and yield for both signal and background. $S_{\text{sig}}$ represents the signal significance for three chosen benchmark points with decay lengths of 1 cm, 50 cm, and 100 cm, and $M_{\chi_2^0}=1600$ GeV and $M_{\chi_1^0}=800$ GeV.}
\label{tab:cutflow}
\end{table}

We generate 0.5 million events for each LLP benchmark. At L1, where we select events with lepton triggers, displaced calo-jet trigger and track-$H_T$ trigger, LLP events are selected with more than 90\% efficiency, with efficiency decreasing as decay length increases. Further, we select events with $H_T >$ 500 GeV. Since QCD events are generated $H_T^{Gen} >$ 500 GeV, most QCD events pass this cut. For CBA, events with at least one secondary vertex with at least six displaced tracks with $M_{DV} >$ 20 GeV are selected. Events with LLPs having smaller decay lengths are mostly selected, while signal efficiency decreases with increasing decay length, with efficiency decreasing to less than 50\% for decay length above 200 cm. We find no background events passing the criteria mentioned above. The instrumental background is also handled since we require $M_{DV} >$ 20 GeV. Next, we select jets out of jets selected at L1 after applying suitable cut on signal probability of jets from MVA-1 and MVA-2 separately where XGBOOST trained model is applied on L1 jets. Those events are selected where we find at least one jet passing the selection criteria on signal probability.

MVA-1 and MVA-2 surpass CBA in identifying LLPs with 200 cm decay length. Remarkably, MVA-2 outperforms CBA notably for $c\tau=$ 200 cm. The combined usage of MVA-1 and MVA-2 demonstrates significantly better signal efficiency than CBA alone, emphasizing the importance of timing information when searching for LLPs with long lifetimes. However, as anticipated, MVA-1 and MVA-2 exhibit poor performance for the LLP benchmark with $c\tau=$ 1 cm compared to CBA. In Table \ref{tab:cutflow}, We also show the yield and signal significance for the three benchmark points for the wino-like chargino-neutralino pair production at HL-LHC, considering an integrated luminosity of $\mathcal{L}=$ 3000 $\mathrm{fb^{-1}}$. We obtain the signal significance of around 9$\sigma$ for all three decay lengths. Remarkably, signal significance does not degrade with decay length, which is attributed to increased analysis sensitivity to the LLPs with higher decay length, thanks to the inclusion of the timing information in the analysis.

\begin{figure}[hbt!]
    \centering
    \includegraphics[scale=0.37]{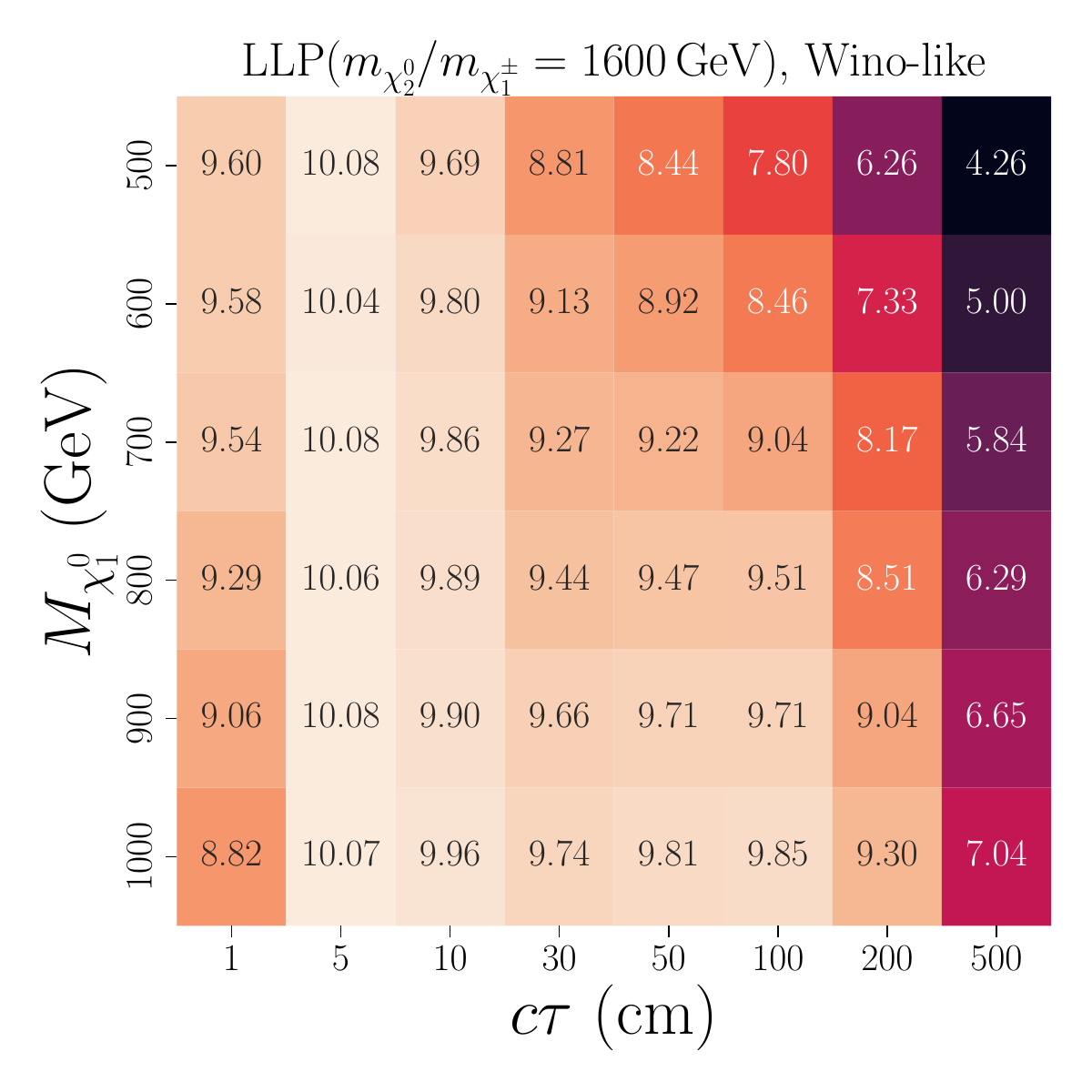}~~
    \includegraphics[scale=0.37]{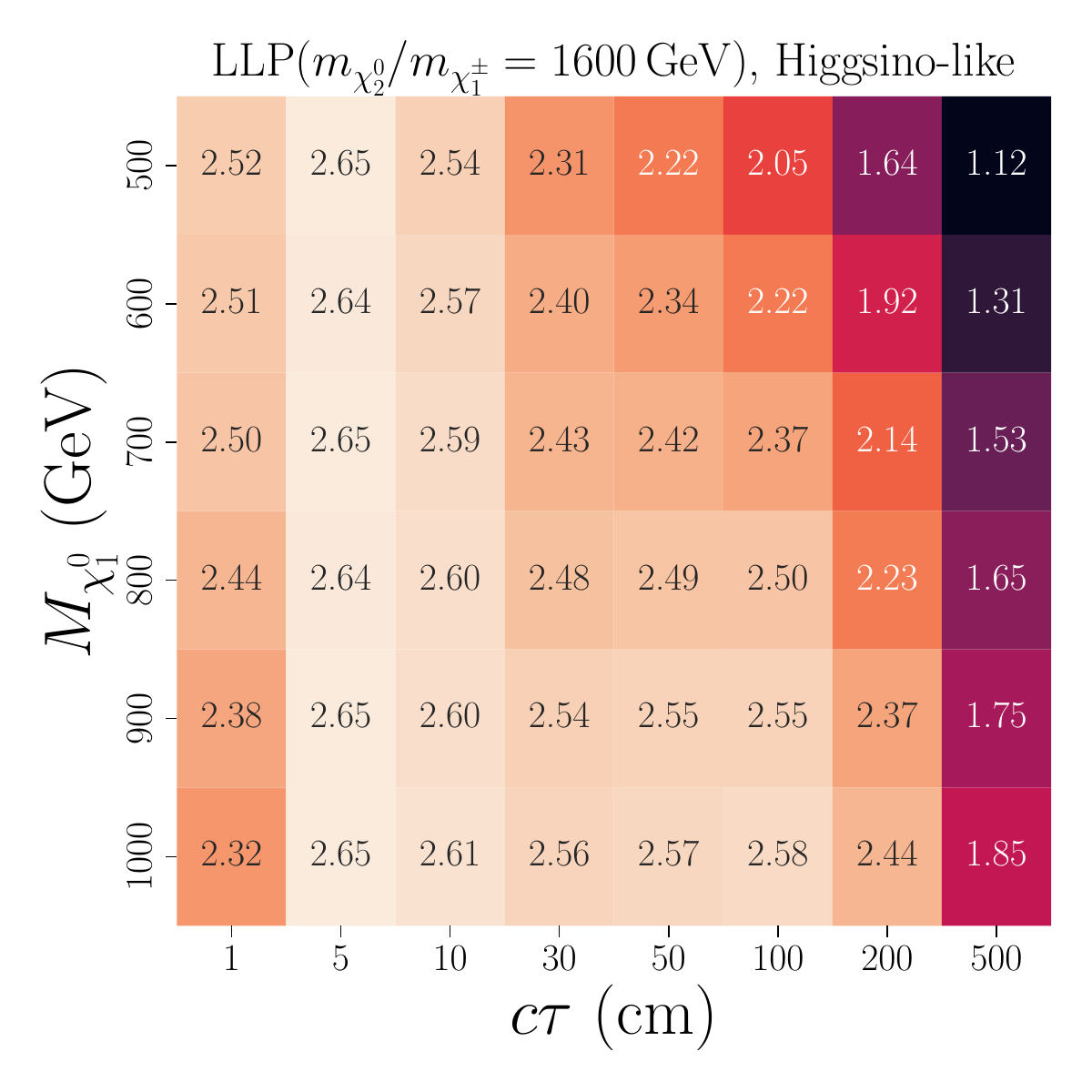}
   
    \caption{Signal significance for LLP benchmark points with $M_{\chi_2^0}=1600$ GeV and $M_{\chi_1^0}$ varying from 500 GeV to 1000 GeV while decay length varies from 1 cm to 500 cm for wino and higgsino-like chargino-neutralino pair production. }
    \label{fig:sig_grid}
\end{figure}

We extend the analysis by calculating signal significance for a set of LLP benchmark points following the similar procedure as described above and in Section \ref{subsec:cba}, \ref{subsec:mva-1}, and \ref{subsec:mva-2}. In Figure \ref{fig:sig_grid}, we present the signal significance for numerous LLP benchmark points for wino and higgsino-like $M_{\chi_2^0}$/ $M_{\chi_1^0}$ pair production scenario where $M_{\chi_2^0}=1600$ GeV and $M_{\chi_1^0}$ varies from 500 GeV to 1000 GeV with decay length varying from 1 cm to 500 cm in the form of a grid. As mentioned earlier, We train three different XGBOOST models for three different decay lengths, namely 1 cm, 50 cm and 200 cm, with $M_{\chi_2^0}=1600$ and $M_{\chi_1^0}=800$. The model trained with LLP benchmark with a decay length of 1 cm is applied on LLPs with decay lengths varying between 1 cm and 5 cm. The LLP model trained with a decay length of 50 cm is applied on LLPs with decay lengths between 10 cm and 100 cm, while the LLP model trained with a decay length of 200 cm is reserved for LLPs with very high decay lengths greater than 200 cm. 

From Figure \ref{fig:sig_grid}, we observe a general trend: the signal significance tends to decrease as the decay length of LLP increases. This results from fewer LLPs decay within the tracker and calorimeter volumes as the decay length of the LLP increases. Moreover, the signal significance decreases with a decrease in the LLP mass. 

For wino-like chargino-neutralino pair production, a maximum signal significance of approximately 10 is observed for LLPs with a decay length of 5 cm across all mass points. At the smallest decay length of 1 cm, the signal significance ranges from 8.82 (for $M_{\chi_1^0}=1000$ GeV) to 9.60 (for $M_{\chi_1^0}=500$ GeV). As the decay length increases to the maximum of 500 cm, the signal significance decreases to a range of 7.04 (for $M_{\chi_1^0}=1000$ GeV) and 4.26 (for $M_{\chi_1^0}=500$ GeV). For $M_{\chi_1^0}=800$ GeV, the signal significance starts at 9.29 for LLP with a 1 cm decay length and drops to 6.29 at a 500 cm decay length. This decrease in signal significance with increasing decay length is consistent across all mass points. Regarding probing and discovery potential at HL-LHC, all mass points exhibit a signal significance greater than two across all decay lengths, implying the potential for probing at the HL-LHC. For discovery potential at the HL-LHC, defined as a signal significance greater than 5, our analysis suggests that all mass points maintain discovery potential up to a decay length of 500 cm except for $M_{\chi_1^0}=500$ GeV at 500 cm decay length.

For higgsino-like chargino-neutralino pair production with a relatively smaller cross-section than a wino-like signature, a maximum signal significance of approximately 2.6 is observed for LLP benchmark points with decay lengths greater than 5 cm and 10 cm across all LLP mass points. At a decay length of 1 cm, the signal significance varies from 2.52 (at $M_{\chi_1^0}=500$ GeV) to 2.32 (at $M_{\chi_1^0}=1000$ GeV). As the decay length extends to 500 cm, the signal significance decreases, with values ranging from 1.85 (at $M_{\chi_1^0}=1000$ GeV) to 1.12 (at $M_{\chi_1^0}=500$ GeV). When considering the probing potential at the HL-LHC, it is important to note that all mass points maintain a signal significance greater than 2 for decay lengths up to 200 cm, except for the LLPs with 500 and 600 GeV mass at a decay length of 200 cm. It is worth mentioning that signal significance increases with the mass of the LLPs. Thus, LLPs with a mass greater than 1000 GeV and a decay length of 500 cm have the potential to be probed at HL-LHC.

\begin{figure}[hbt!]
    \centering
    \includegraphics[scale=0.37]{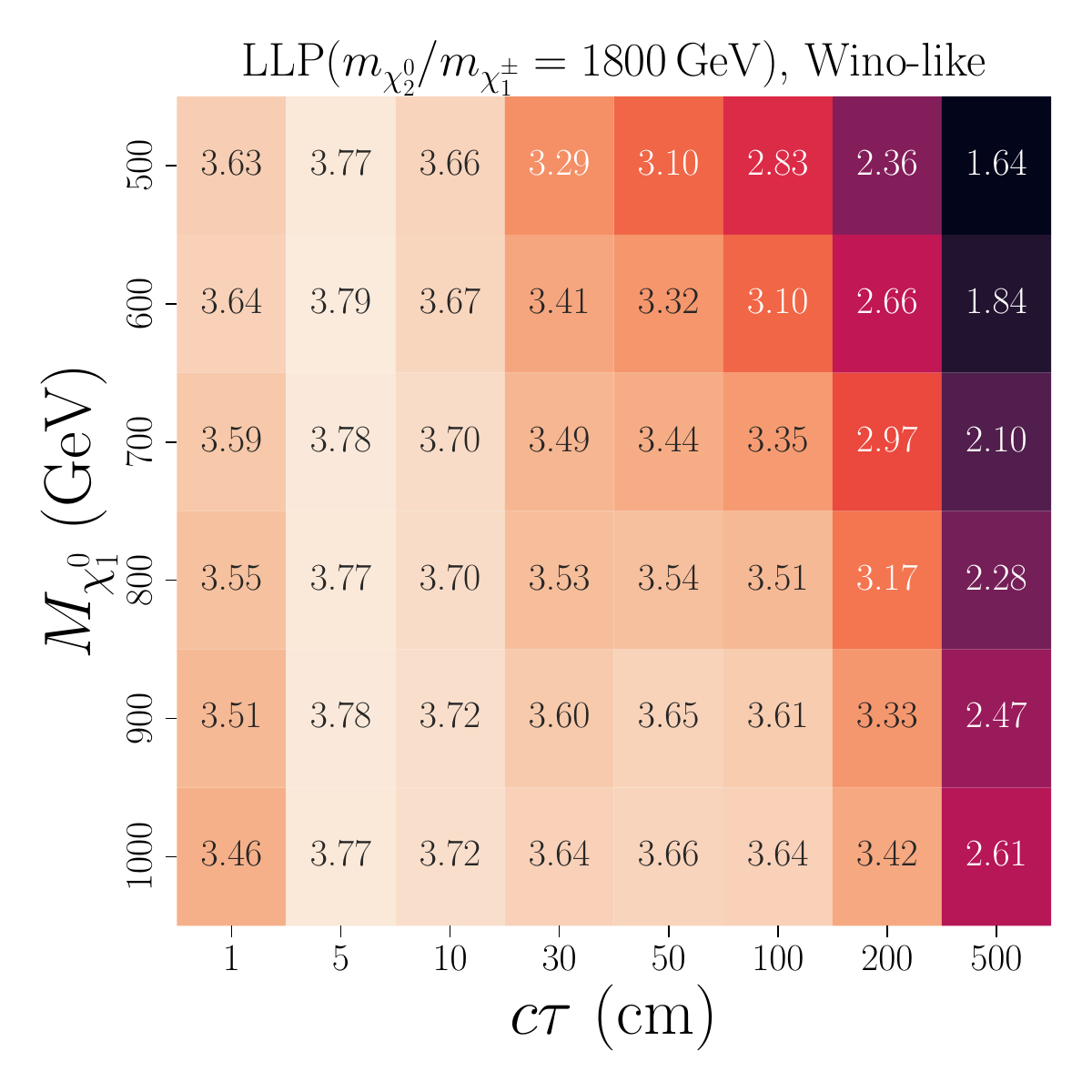}~~
    \includegraphics[scale=0.37]{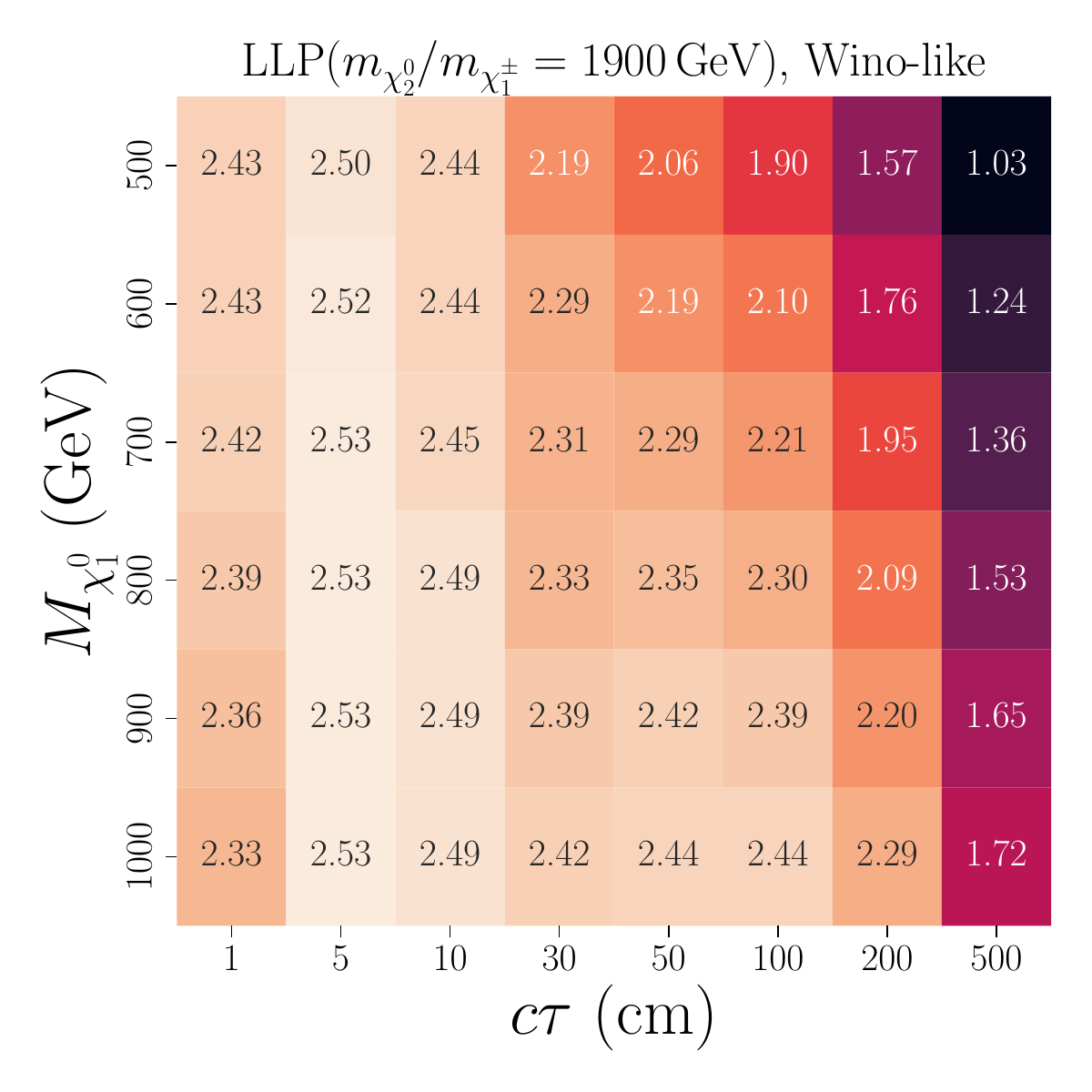}
   
    \caption{Signal significance for LLP benchmark points with $M_{\chi_2^0}=1800$ GeV and 1900 GeV and $M_{\chi_1^0}$ varying from 500 GeV to 1000 GeV while decay length varies from 1 cm to 500 cm for wino-like chargino-neutralino pair production. }
    \label{fig:sig_grid_extra}
\end{figure}

We also present signal significance for wino-like chargino-neutralino pair production for $M_{\chi_2^0}=1800$ GeV and 1900 GeV as shown in Figure \ref{fig:sig_grid_extra}.

For $M_{\chi_2^0}=1800$ GeV, we observe a maximum signal significance of $\approx$3.8 at $M_{\chi_1^0}=600$ GeV with a 5 cm decay length. LLPs with mass $M_{\chi_1^0}>600$ GeV maintain a signal significance over two at a decay length of 500 cm, indicating the potential for probing at the HL-LHC. For $M_{\chi_2^0}=1900$ GeV, signal significance generally decreases with increased decay lengths, but higher mass points retain better values, suggesting a stronger HL-LHC probing potential. Particularly at the decay length of 200 cm, all mass points from $M_{\chi_1^0}=800$ GeV and above maintain signal significance above the threshold of 2. For LLP with $M_{\chi_1^0}=1000$ GeV and 500 cm decay length, a signal significance of 1.72 is observed, which is close to the probing threshold. Furthermore, as the mass of the LLPs at the decay length of 500 cm increases, the signal significance is expected to rise further. Therefore, LLPs with masses greater than 1000 GeV could be probed at HL-LHC.

\section{Summary and conclusion}
\label{sec:summary}
The exploration of supersymmetry (SUSY) continues to be crucial in investigating physics beyond the Standard Model, driven by strong theoretical and phenomenological motivations. Although both R-parity conserving and violating scenarios of SUSY have been extensively studied using prompt physics signatures, there is a scarcity of realistic phenomenological studies targeting the search for SUSY via exotic displaced signatures in R-parity violating (RPV) SUSY, especially in the context of HL-LHC. In this work, we particularly analyze the pair production of electroweakinos, $\chi_2^0$ and $\chi_1^{\pm}$, and their decay into Higgs boson and W boson, respectively, along with $\chi_1^0$. The $\chi_1^{0}$ then undergo further decay to light quarks, facilitated by small values of the RPV couplings $\lambda^{''}$, resulting in $\chi_1^{0}$  with longer lifetimes.

In order to efficiently select events at the Level-1 trigger level, we have used three triggers: Track-$H_T$, Displaced Calo-Jet, and Single TkIsoLepton. The first two triggers are specifically designed for displaced searches. Our analysis shows that the Displaced Calo-Jet trigger is highly effective in selecting long-lved particle (LLP) events where LLP has a longer lifetime, while the Track-$H_T$ trigger is primarily efficient in selecting LLP events with smaller decay lengths. By combining above these three triggers, we demonstrate the ability to effectively select LLP events across a wide range of decay lengths, ranging from very small to very high, with a high level of efficiency. This highlights the complementary nature of these triggers in capturing LLP signatures with varying decay lengths, and underscores their effectiveness in our study. In the following step, we construct several physics variables by utilizing information from the tracker, MTD, and calorimeters. The analysis is subdivided into three parts, namely cut-based analysis (CBA), multivariate analysis-1 (MVA-1), and multivariate analysis-2 (MVA-2). The cut-based analysis incorporates displaced vertex information, while MVA-1 and MVA-2 employ timing information from MTD and ECAL, respectively. Our findings indicate that LLPs with shorter decay lengths can be effectively searched using the cut-based analysis. However, for LLPs with longer decay lengths, where displaced vertex information alone may not be sufficient, timing-based analyses such as MVA-1 and MVA-2 provide effective selection methods. These results contribute to the understanding of best approaches for identifying LLPs in different decay length scenarios, considering the limitations of displaced vertex information and the potential of timing-based analyses in the context of this study. 

Finally, we calculate the signal significance for LLPs in different benchmark scenarios. We vary the mass of LLPs from 500 GeV to 1000 GeV and the decay length from 1 cm to 500 cm for both wino-like and higgsino-like electroweakino pair production scenarios, with a degenerate chargino/neutralino mass, $M_{\chi_2^0}/M_{\chi_1^{\pm}}$ = 1600 GeV. Our results show that LLPs in the wino-like chargino/neutralino pair production scenario, for all benchmark points discussed, have the potential to be probed at the HL-LHC with signal significance greater than or equal to 5$\sigma$ for all LLP masses except for LLP with mass 500 GeV at 500 cm decay length where signal significance is less than 5 but greater than 2. However, the significance decreases for the higgsino-like scenario. Nonetheless, the majority of the benchmark points exhibit signal significance greater than 2$\sigma$ except for LLPs at 500 cm decay length and LLPs with mass  $\leq$600 GeV at 200 cm decay length, suggesting that they can be probed at the HL-LHC. In comparison, the ATLAS study \cite{ATLAS:2023oti} which examines the pair production of electroweakinos in four channels in a pure higgsino state, using the processes $pp \rightarrow \chi_1^{\pm}\chi_2^{0}$, $\chi_2^{0}\chi_1^{0}$, $\chi_1^{+}\chi_1^{-}$, and $\chi_1^{\pm}\chi_1^{0}$ at 13 TeV, rules out electroweakinos with masses below roughly 1250 GeV for a decay length of 200 cm. Our analysis, focusing only on the $\chi_1^{\pm}\chi_2^{0}$ production channel, projects the exclusion mass limit for electroweakinos to 1600 GeV at the same decay length. 

We also calculate the signal significance for the heavier electroweakinos with $M_{\chi_2^0}/M_{\chi_1^{\pm}}$ of 1800 and 1900 GeV for wino-like LLP signatures. For  $M_{\chi_2^0}/M_{\chi_1^{\pm}}=$1800 GeV, all mass points, except for $M_{\chi_1^0}\leq$ 600 GeV, retain a signal significance above 2 across all decay lengths. For $M_{\chi_2^0}=1900$ GeV, Despite the general decrease in signal significance with increased decay lengths, higher mass points sustain stronger values, thereby indicating their probing potential at HL-LHC. Particularly, all mass points from $M_{\chi_1^0}=800$ GeV and higher maintain a signal significance above 2 at a decay length of 200 cm indicating their probing potential at HL-LHC.

\section*{Acknowledgement}
BB acknowledges the support provided by the MATRICS Grant(MTR/2022/000264) of the Science and Engineering Research Board (SERB), Government of India.
\clearpage
\newpage
\appendix
\section*{Appendix}
\section{Correlation matrix for ECAL timing variables}
\label{app:corr}

\begin{figure}[hbt!]
    \centering
    \includegraphics[scale=0.75]{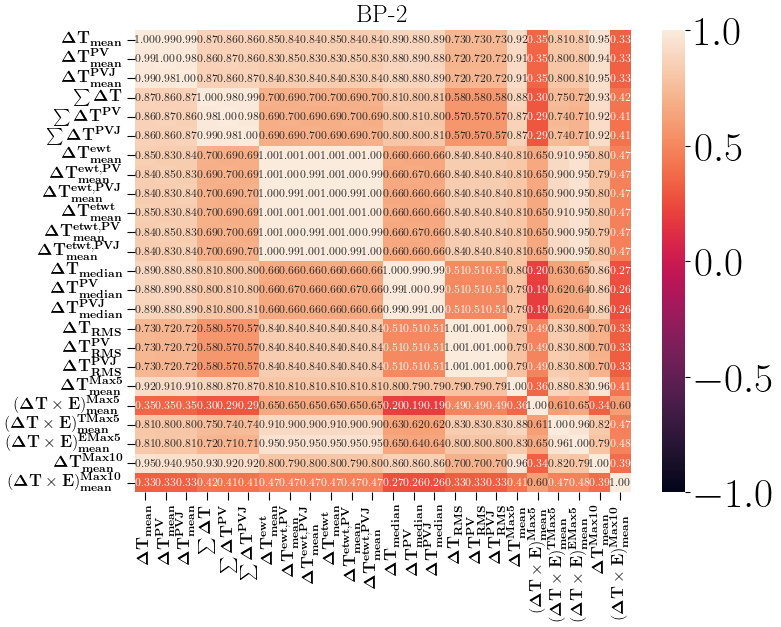}~~
\caption{Correlation between different timing variables constructed using information from ECAL for signal (BP-2). }
    \label{fig:corr_llp}
\end{figure}

\begin{figure}[hbt!]
    \centering
    \includegraphics[scale=0.75]{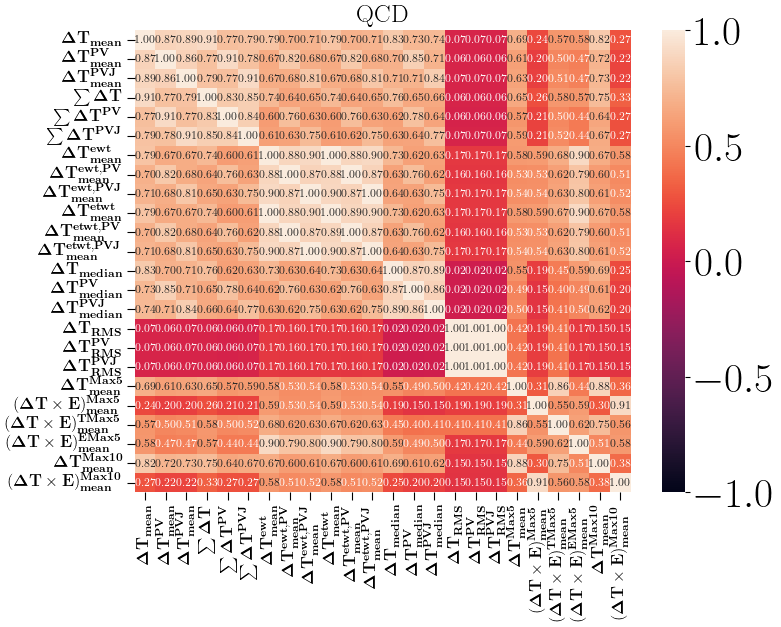}~~
\caption{Correlation between different timing variables constructed using information from ECAL for background (QCD). }
    \label{fig:corr_qcd}
\end{figure}
\newpage
\bibliographystyle{JHEP.bst}
\providecommand{\href}[2]{#2}\begingroup\raggedright\endgroup
\end{document}